\documentclass[sigconf, nonacm]{acmart}
\usepackage{svg}
\usepackage{booktabs} 

\usepackage{exii-macros}

\usepackage{booktabs}
\usepackage{array}
\usepackage{subcaption}
\usepackage{fontawesome}
\usepackage{multirow}
\usepackage{graphicx}
\usepackage{wrapfig}
\usepackage{lipsum}
\usepackage{makecell}

\definecolor{codecolor}{RGB}{239,239,237}
\definecolor{codetextcolor}{RGB}{165,0,0}

\newcommand{\inlinecode}[1]{%
  \colorbox{codecolor}{\textcolor{codetextcolor}{\small \texttt{#1}}}%
}

\makeatletter
\g@addto@macro\normalsize{%
  \setlength\abovedisplayshortskip{-9pt}
  \setlength\belowdisplayshortskip{3pt}
}
\makeatother

\begin{document}

\tolerance=400 

%
\title[Code Shaping]{Code Shaping: Iterative Code Editing with Free-form AI-Interpreted Sketching}
\newcommand{\sys}[0]{\f{Code Shaping}}
\newcommand{\baseline}[0]{\textit{Baseline}}



\author{Ryan Yen}
\orcid{0001-8212-4100}

\affiliation{%
  \institution{School of Computer Science, University of Waterloo}
  \country{}
}
\affiliation{%
  \institution{CSAIL, MIT}
  \streetaddress{77 Massachusetts Ave}
  \country{}
}
\email{ryanyen2@mit.edu}

\author{Jian Zhao}
\orcid{0002-7761-6351}

\affiliation{%
  \institution{School of Computer Science, University of Waterloo}
  \country{}
}
\email{jianzhao@uwaterloo.ca}

\author{Daniel Vogel}
\orcid{0000-0001-7620-0541}
\affiliation{%
  \institution{School of Computer Science, University of Waterloo}
  \country{}
}
\email{dvogel@uwaterloo.ca}

\renewcommand{\shortauthors}{Yen et al.}

\begin{abstract}

We introduce the concept of code shaping, an interaction paradigm for editing code using free-form sketch annotations directly on top of the code and console output. To evaluate this concept, we conducted a three-stage design study with 18 different programmers to investigate how sketches can communicate intended code edits to an AI model for interpretation and execution. The results show how different sketches are used, the strategies programmers employ during iterative interactions with AI interpretations, and interaction design principles that support the reconciliation between the code editor and sketches. Finally, we demonstrate the practical application of the code shaping concept with two use case scenarios, illustrating design implications from the study.




\end{abstract}

%
%



\begin{teaserfigure}
\centering
  \includegraphics[width=\linewidth]{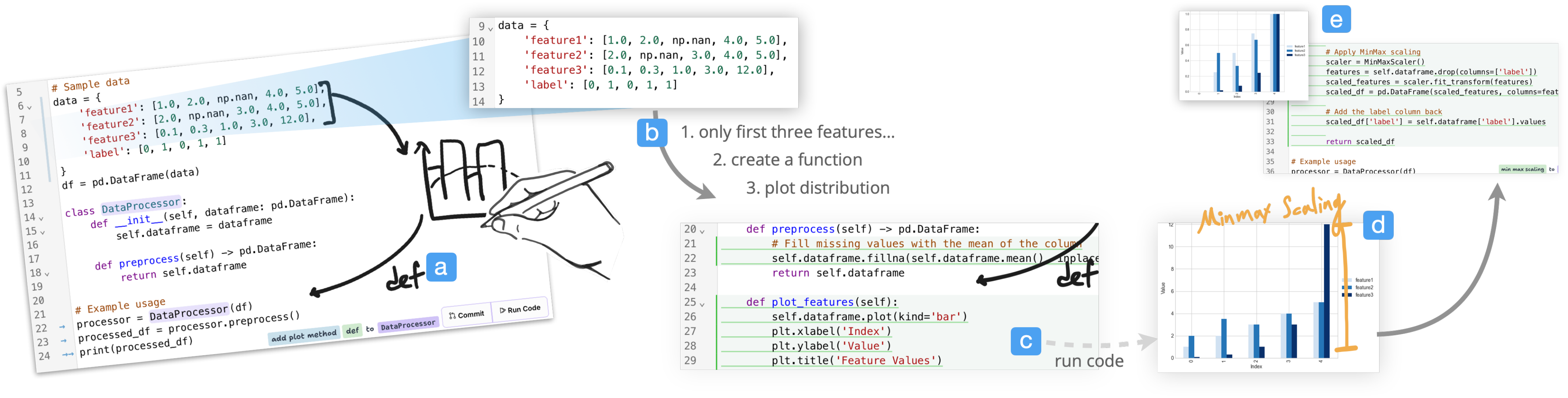}
  \caption{Code shaping usage example: (a) a programmer draws an arrow from a few lines of code defining data attributes to a sketch of a bar chart in whitespace near the code, then they add another arrow back to a different code location and annotate the arrow with `def'; (b) an AI model uses the code and the overlaid sketches to insert a new function to plot that data; (c) the programmer reviews the edits interpreted by the model, then they run the program; (d) the code outputs a rendered plot, the programmer sketches on top of it to indicate it should use min-max scaling; (e) the model examines the new sketches and modifies the code to implement scaling.}  
  \Description[Code shaping process with AI assistance.]{The figure illustrates an interactive code shaping process that integrates programmer input and AI assistance. It begins with (a) a Python code editor displaying a data preprocessing script and a user adding handwritten annotations and arrows linking the code to a sketched bar chart. A label 'def' is drawn to indicate the function definition process. Then, in step (b), the AI interprets these annotations, inserting a plot_features function into the code editor. In step (c), the programmer reviews and runs the updated code, and in (d), the output is displayed as a bar chart with handwritten annotations indicating "Min-max scaling." Finally, step (e) shows the AI modifying the code further to include scaling functionality. The process demonstrates an iterative collaboration between the programmer and AI, highlighted by handwritten notes and diagrams across stages.}
  \label{fig:teaser}
\end{teaserfigure}


\maketitle



\section{Introduction}
In programming tasks, text is not always the primary medium for expressing ideas \cite{latoza_maintaining_2006}. Programmers often turn to sketching on whiteboards and paper to externalize thoughts and concepts \cite{cherubini_lets_2007, 6922572, 6065018}. This includes tasks like designing program structure, working out algorithms, and planning code edits~\cite{cherubini_lets_2007, 10.1145/1879211.1879217, sutherland_investigation_2017}.
The informal nature of sketching helps untangle complex tasks, represent abstract ideas, and requires less cognitive effort to comprehend \cite{cherubini_lets_2007, tversky2002sketches, goel1995sketches}.

Prior research has explored programming-by-example systems that transform sketches~\cite{10.1145/22627.22349}, such as diagrams~\cite{10.1145/1281500.1281546}, mathematical symbols~\cite{li2008algosketch, 10.1145/3411764.3445460}, and user interfaces~\cite{tldraw, 910894, microsoft_sketch2code}, into functional programs.
However, these systems often target non-programmers, with the generated code typically hidden or not intended for direct editing.
For programmers, another line of research has enhanced current integrated development environments (IDE) with sketch-based annotation features from the engineering perspective to support note-taking~\cite{sutherland2015observational, 10.1145/1324892.1324935}, facilitate collaboration~\cite{lichtschlag2014codegraffiti}, and aid in planning future code edits~\cite{samuelsson2020eliciting}.
Despite these advancements, sketching and code editing are still largely treated as separate activities in the software development process.

This division stems from the traditional view of programming as primarily text-based~\cite{arawjo_write_2020}, with sketching seen as an auxiliary tool.
Programmers must switch contexts between sketching and coding, potentially losing insights during the translation from visual ideas to code modifications~\cite{parnin2006building, 10.1145/1879211.1879217, bff9b250-7640-39e2-8f34-329fd1552822}. This challenge is exacerbated by the non-linear and dynamic nature of programming, where code is frequently revisited and revised in response to evolving requirements and new discoveries.
Hence, sketches have been primarily considered as a static external representation of the programmer's thoughts instead of ways to interact with code~\cite{sutherland2015observational, 10.1145/1324892.1324935, 1698771}.

To address this separation, we propose a sketch-based editing approach where a \textit{programmer draws free-form annotations on and around the code to iteratively guide an AI model in modifying code structure, flow, and syntax}: a concept we call \textit{code shaping}. For example, to insert a function to visualize data, a programmer can circle lines of code related to data attributes, draw an arrow to a sketch of a graph, then draw another arrow with the word ``def'' back to a line of code to insert the function (\autoref{fig:teaser}a,b,c). Further iterations of sketching can revise the function name or specify additional data processing steps (\autoref{fig:teaser}d,e). This approach tightly integrates free-form sketching with realtime code editing both visually and operationally, providing programmers with an alternative modality to express modifications.
This approach allows programmers to encapsulate their expectations for the program's functionality and link these sketches directly to syntactic code. However, challenges such as model interpretation errors due to the inherent ambiguity of sketches~\cite{10.1145/1281500.1281527, 10.1145/237091.237119} and the fundamental differences between sketching and coding modalities require further design exploration.

We adopted a user-centered design process with 18 programmers using a prototype system probe that implements the code shaping concept. Our findings reveal the types of sketches programmers created, their strategies for correcting AI model errors, and design implications for bridging the conceptual gap between the canvas where sketches are made, the textual code representation, and the AI models. We demonstrate these design implications with two real-world use cases: a productivity break using a tablet and pair programming at a whiteboard.
The contribution of this research is not to claim that code shaping is superior to other interaction paradigms, such as typing, but to establish it as a viable alternative that empowers programmers to iteratively express and refine their code edits through free-form sketches. 

\section{Related Work}
Historically, the practice of \textit{writing code} has evolved significantly alongside the development of different tools and technologies. Early methods relied on handwritten and drawn notations, reflecting the material and cultural contexts of their time~\cite{arawjo_write_2020}. 
The advent of the typewriter marked a pivotal shift, standardizing typed input as the dominant mode for programming.
However, multiple explorations into alternative, keyboard-less methods of code manipulation have been conducted. Research has investigated the use of gestures~\cite{murphy2011restructuring}, touch inputs~\cite{tillmann2012touchdevelop, raab_refactorpad_2013, 10.1145/1879211.1879217}, and \rev{voice- or speech-based input~\cite{arnold2000programming, desilets2001voicegrip} for programming. These studies demonstrate a consistent effort to move beyond traditional text-based coding by leveraging different interaction modalities to make programming more accessible.}

These studies demonstrate a consistent effort to move beyond traditional text-based coding by leveraging different interaction modalities to simplify and enhance the coding experience.
\rev{The advent of large language models (LLMs) marks another paradigm shift in the way code is written. With capabilities for code generation and completion from natural language, LLMs have made the long-envisioned concept of literate, unstructured, and natural programming more feasible~\cite{knuth1984literate, bobrow1964natural, weizenbaum1966eliza}. The use of LLM-driven code assistants is transforming programming workflows, as developers increasingly transition from writing code manually to critically evaluating and refining AI-generated code~\cite{barke_grounded_2022, mozannar_reading_2022}.}
While the keyboard remains a central tool, advancements in computer vision and speech recognition are expanding the possibilities of programming, allowing for diverse and multimodal forms of code input and manipulation~\cite{pollock2024designing, horowitz2023live}.
Among these modalities, sketching has attracted significant attention as a flexible and expressive method for generating code.

\subsection{Generate Code from Sketches}
\label{sec:sketch-to-code}
Prior work has explored the transformation of sketches into code to facilitate rapid prototyping and early-stage design. Tools like SILK~\cite{landay1995interactive, 910894} enable designers to sketch UI elements electronically, turning them into interactive prototypes, thereby supporting flexible sketching and demonstrating the effectiveness of sketch-based methods for generating functional UIs. DENIM~\cite{10.1145/332040.332486} extends this approach by offering a zoomable user interface that supports web design sketches across multiple levels of detail, from high-level site maps to specific page elements.
Other tools, such as Eve~\cite{10.1145/3290607.3312994}, provide a comprehensive sketch-based prototyping workbench that facilitates transitions between low, medium, and high-fidelity prototypes, ultimately generating executable code. More recent approaches, like pix2code~\cite{10.1145/3220134.3220135} and Microsoft's Sketch2Code~\cite{microsoft_sketch2code}, leverage deep learning and computer vision techniques to convert GUI sketches into code for multiple platforms. Although these tools demonstrate the utility of sketches in generating code, they primarily focus on sketching the program output and transforming them into code, rather than using sketches as a direct manipulation method for editing code.

Further, these sketches often exist in separate mediums from the code, and sometimes the code might not even be shown alongside them~\cite{tldraw}. 
This separation makes direct visual-to-code mappings challenging~\cite{cherubini_lets_2007} since code is inherently abstract without definitive representation.
This often results in sketches being transient, as they are attempts to translate fluid visual representations into the structured syntax required by code~\cite{4782972}. The temporary nature of these sketches highlights the difficulty in maintaining a clear mapping between visual sketches and syntactic code structures. Arawjo et al.~\cite{arawjo_notational_2022} introduced notational programming, which integrates small canvases containing handwritten notations within code cells of computational notebooks, showing an initial effort to merge sketches and code. However, this approach maintains only an implicit connection between code and sketches, limiting explicit linkage between handwritten symbols and their textual equivalents.
In contrast, our proposed concept, \textit{code shaping}, goes beyond both notational programming and programming-by-example approaches\cite{10.1145/22627.22349}. It enhances the linkage between sketches and code by allowing programmers to sketch directly on and around the code, resembling a visual programming language. This enables visual planning and referencing of future edits, fostering a more direct and dynamic interaction between sketches and actual code. 




\subsection{Annotating and Planning Code with Sketches}
Programmers often use sketches, highlights, and external notes to annotate code for better comprehension, resource tracking, progress monitoring, and peer communication~\cite{maalej_comprehension_2014, sutherland_investigation_2017}. Several systems have been developed to support these annotation practices. For instance, Synectic IDE~\cite{synectic} facilitates linking and annotating code files to assist in programming tasks. However, for annotations to be effective, they should be integrated directly into the code editor or positioned close to the code to help programmers maintain their workflow~\cite{parnin_evaluating_2010}.
Annotations lacking context from surrounding code can hinder understanding of their implications for future edits~\cite{maalej_comprehension_2014}. Systems like Catseye have addressed this issue by enabling programmers to add contextually linked annotations alongside the code editor, serving as a note-taking tool~\cite{horvath_using_2022}.
However, Catseye's annotations are limited to typed textual notes linked to code snippets, lacking the flexibility offered by freeform sketches.
Recent research in software engineering has focused on developing integrated development environments (IDEs) that allow programmers to sketch directly on the code editor for note-taking, such as CodeAnnotator~\cite{10.1145/1324892.1324935} and CodeGraffiti~\cite{10.1145/1866218.1866260}.
However, similar to the sketch-to-code approaches discussed in \autoref{sec:sketch-to-code}, these sketches primarily serve as static externalizations of users' thoughts rather than interactive mediums for code manipulation.
Consequently, sketches and programs remain separate modalities with distinct affordances. This limitation restricts the practical application of sketching on code to scenarios involving code comprehension or collaborative discussions.

However, integrating sketch-based annotations for planning code edits with subsequent code modifications presents significant challenges due to the inherent nature of the program.
First, the dynamic and interdependent nature of code means changes in one part can have cascading effects across the entire codebase, complicating the predictive power of annotations. Second, the code's non-linear nature, where functions, variables, and classes interact in complex, non-sequential ways, requires annotations to account for these intricate relationships. Third, the rigid syntax and structure of code demand precise and well-integrated annotations, unlike the more flexible and informal notes used in text editing.
Despite these challenges, using sketched annotations for planning code edits could play a significant role in software development~\cite{cherubini_lets_2007}.
Samuelsson et al. investigated common sketches for standard code editor commands, such as inserting or searching code~\cite{samuelsson_towards_2023}. However, they only considered using sketches as replacements for common IDE commands, similar to previous research using gestures to replace keyboard shortcuts~\cite{raab_refactorpad_2013}.
Our research, in contrast, explores the potential of transforming these annotations into actionable commands for \textit{code editing}, allowing programmers to make annotations and edits without constant context switching between code and external spaces like canvases or paper.

\begin{figure*}
    \centering
    \includegraphics[width=\linewidth]{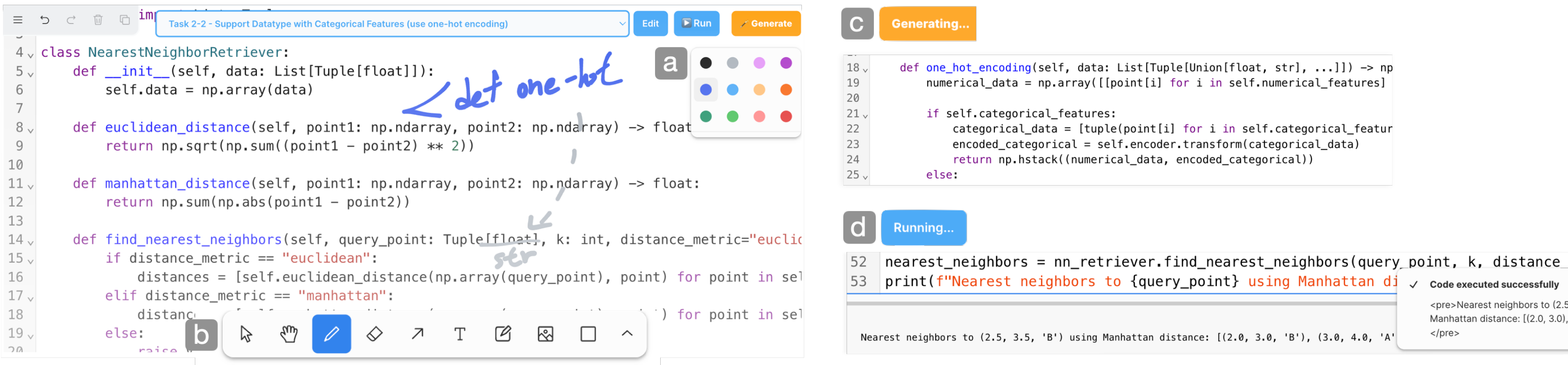}
    \caption{Interface design from the first stage. (a) Pen tool with color options for code annotation; (b) Canvas tools including select, pan, pen, eraser, and other common shapes; (c) AI-powered ``Generate'' button for translating sketches to code edits; (d) ``Run'' button executes Python code and displays output in the console.}
    \label{fig:first-interface}
    \Description[Coding interface with annotation and execution features.]{The figure presents a coding interface design with annotation and execution features. The main area displays Python code implementing a NearestNeighborRetriever class, featuring methods for Euclidean and Manhattan distance calculations and a nearest neighbor search. (a) Blue handwritten annotations, such as "def one-hot," are visible directly on the code. (b) A toolbar below the interface offers tools like select, pan, pen, eraser, and shape drawing, along with a color palette for annotations displayed as colorful dots to the right. (c) An "AI-powered Generate button" appears near the top, used for translating user sketches into code edits. (d) The "Run" button executes the code and shows the results in the console, displayed in the bottom right. The console output includes nearest neighbor results based on Manhattan distance. This setup demonstrates an interactive workflow where users annotate, generate, and execute code within an accessible interface.}
\end{figure*}

\section{Code Shaping}
The code shaping concept \emph{enables programmers to edit code using freeform sketches directly on or around the code}. This approach includes three core elements: a sketching canvas, a responsive code editor, and an AI that interprets sketches to generate code edits.

In a code shaping session, programmers sketch their intended modifications on an invisible canvas overlaid on the code. These sketches can include arrows pointing to specific lines, pseudocode defining a function's structure, and annotations indicating desired changes. The sketches can interact with any part of the code and output in the console and graphical view.
Once sketches are made, users can press a button to prompt AI to interpret their sketches along with the code. 
If the resulting code does not match the programmer's intent, they can refine their sketches, creating an iterative cycle of input and feedback.
This feedback loop allows programmers to use sketches progressively and iteratively to \textit{shape} how the code should be structured, how it should flow, and how it should function, guiding it towards the desired form and functionality.

In the following sections, we describe a series of three design studies (stages) to develop a proof-of-concept system and interface for the core code shaping interactions.
The first stage examined the types of annotations used in code shaping. The second stage focused on exploring model interpretation errors and the strategies programmers employed to address them. The final stage synthesized prior stages' insights, aiming to coordinate the interactions when editing code, iterating with AI, and sketching on the canvas. 

\section{Stage One: Explore Sketches} 
We developed a basic user interface to explore how participants used sketches as actionable commands for code edits. We observed and categorized participants' challenges and sketch types, providing foundational insights for the code shaping system's development in subsequent stages.

\subsection{User Interface}
\begin{wrapfigure}{l}{15mm}
\vspace{-3mm} \includegraphics[]{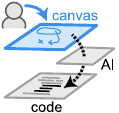}
\end{wrapfigure} 
For this stage, the user interface creates a straightforward way to make free-form sketches in the \textit{canvas layer} to directly generate code edits affecting the \textit{code layer} while keeping the \textit{AI layer} hidden to the user.
The interface supports typical free-form sketching tools, including colour selectors, pens, erasers, and shapes (\autoref{fig:first-interface}a-b). A text tool is available for conventional editing. 
Two-finger panning and zooming navigate the code in the editor to enable sketching at different levels of granularity. A pointer tool can select strokes in the sketches. Pressing a ``Generate'' button uses all annotations on the canvas, or only selected annotations, as parameters for generating edited code (\autoref{fig:first-interface}c). 
The system recognizes free-form annotations on the code editor, utilizing GPT-4o to generate corresponding code. 
We render HTML content from both the code editor and sketches onto separate canvases and embed these canvas content into an SVG. This transformation process includes handling CORS and tainting issues, adding grids to locate annotations, and turning the code editor to grayscale to highlight the sketches.
The system then considers the annotations alongside previous
\rev{version history, including pictorial representations of sketches, code snapshots, conversational context, natural language inputs, and modified code outputs. The stored history was used to contextualize model responses and maintain a comprehensive context of the evolving codebase.}
After the code is generated, a difference algorithm is employed to 
only update the changed sections of the code~\cite{myers1986nd}.
The user can press a ``Run'' button to execute the code, with text or image results shown in the console panel underneath (\autoref{fig:first-interface}d). Programmers can annotate any output on the console or graphical windows as part of their sketches. \rev{The system incorporates these annotated outputs by transforming them into separate canvases, embedding them as SVGs alongside the code editor content, and encoding them for processing.}

\subsection{Participants, Tasks, and Procedure}
We recruited 6 programmers (1 left-handed), aged 23 to 28, with 4 identifying as women and 2 as men.
Participants were recruited through convenience sampling and received \$30 for completing the study. Based on a screening questionnaire, participants had 2-8 years of programming experience in Python and had used ChatGPT or Copilot 3-12 times per week.

We developed three Python coding scenarios, each comprising two tasks that required specific edits to achieve predefined goals. These scenarios spanned different programming paradigms: basic object-oriented programming, functional programming for machine learning, and declarative programming for data engineering. Each task provided participants with starter code requiring modifications in multiple areas. For instance, \f{scenario 2} involved extending a class to handle categorical features in data points, necessitating changes to existing methods for feature encoding and distance calculations. All tasks were pre-tested to ensure that GPT-4o could not immediately generate the correct code.

Participants were assigned 2 out of 3 scenarios that they were most familiar with, as determined by their screening questionnaire. Each scenario consisted of 2 tasks, and participants completed all 4 tasks (2 \f{scenarios} \by 2 \f{tasks} each) within a total of 60 minutes, spanning around 12 to 16 minutes per task. The order of scenarios and tasks was pre-assigned, meaning participants completed all tasks within one scenario before moving on to the next scenario. 
The study was conducted in person using an Apple iPad Air (5th generation, 10.9-inch display) as the primary research tool. An experimenter was present throughout each session to observe and document participant behaviours.
Finally, a semi-structured interview gathered qualitative data on participants' general experience, challenges encountered, and suggestions for system improvement.

\subsection{Data Analysis}
We conducted an inductive thematic analysis of participant-generated sketches. This analysis incorporated observational notes, screen recordings, transcribed think-aloud data, and interview notes. 
Sketches were automatically captured in base64 format each time the generate button was activated, yielding 81 distinct screenshots. 
Of these, 7 were identified as duplicates and subsequently removed from the analysis.
We developed a codebook covering five dimensions: Content (text, code, annotation, freeform), Approach (step-by-step, one-time), CodeReference (parameters, targets), Purpose (functional, procedural), and Form (concrete, abstract). All captured sketches were verified and coded by researchers together. The results and descriptions of each coding category are presented in \autoref{tab:first_code}.



\begin{figure}
    \centering
    \includegraphics[width=.57\linewidth]{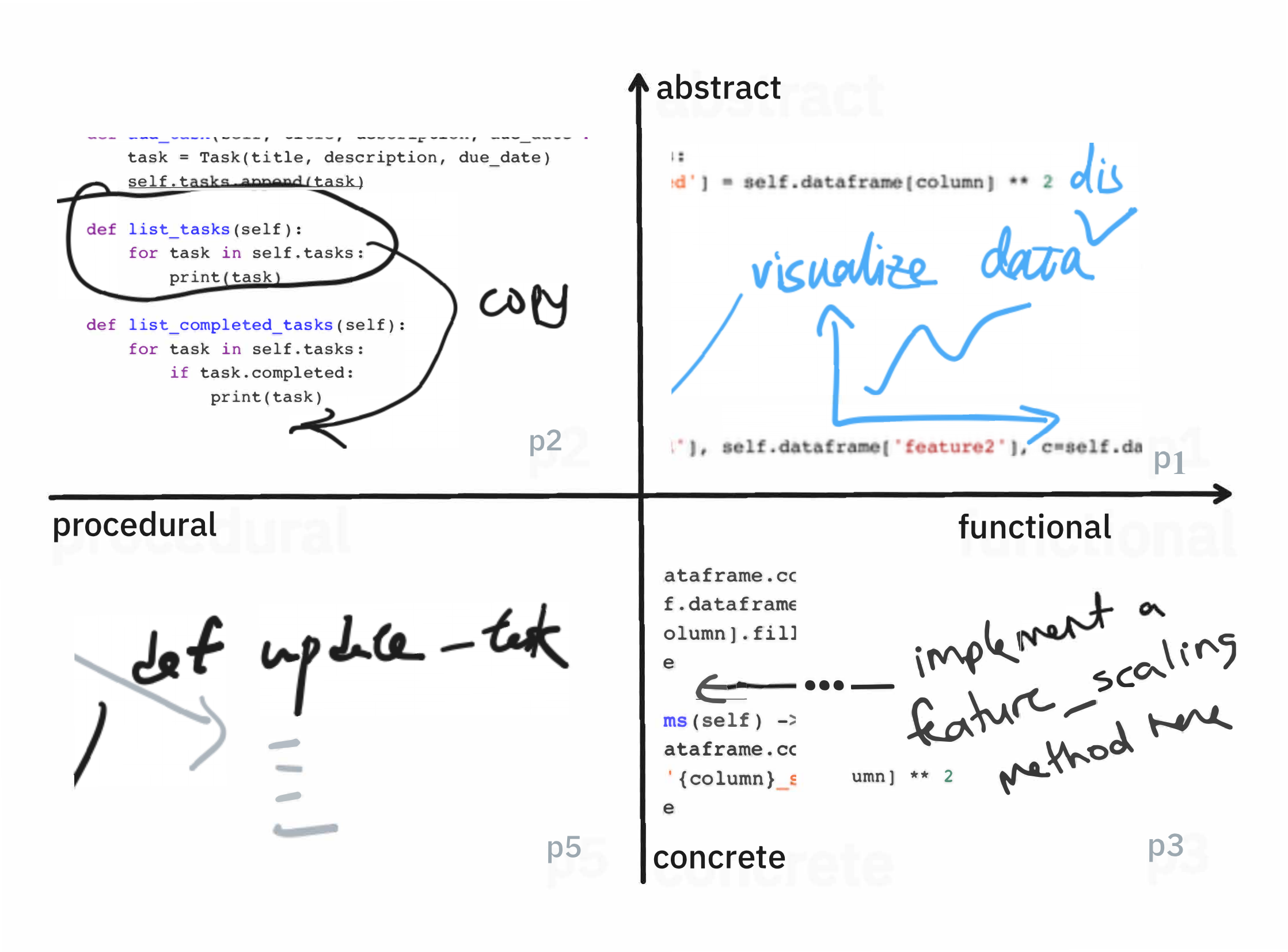}
    \caption{The classification of sketched annotations from participants situated in a quadrant with two spectra, Abstract-Concrete and Procedural-Functional.}
    \Description[Quadrant diagram classifying sketched annotations.]{The figure presents a quadrant diagram categorizing sketched annotations based on two axes: Abstract-Concrete (vertical axis) and Procedural-Functional (horizontal axis).1) The upper-left quadrant (Abstract-Procedural) contains Python code snippets for listing tasks, annotated with handwritten notes such as "list tasks" and arrows connecting functions; 2) The upper-right quadrant (Abstract-Functional) features a hand-drawn graph labeled "visualize data" alongside code suggesting data visualization processes; 3) The lower-left quadrant (Concrete-Procedural) includes a handwritten note "def update_task" with an arrow pointing to a list structure, suggesting task update functionality; 4) The lower-right quadrant (Concrete-Functional) displays a code snippet for manipulating a dataframe column, accompanied by a handwritten comment describing the implementation of a feature scaling method. This diagram illustrates the interplay between abstract and concrete ideas as well as procedural and functional approaches in annotating and conceptualizing code.}
    \label{fig:classification}
\end{figure}

\begin{table*}[]
    \centering
    \small
\begin{tabular}[t]{l|lp{6.1cm}rp{6.2cm}}
\toprule
\textbf{Category} & \textbf{Subcategory} & \textbf{Description} & $N$ & \textbf{Example} \\
\midrule
\multirow[t]{4}{*}{Content} 
    & Text        & Written text or natural language instruction & 11 & ``impute missing value [...]'' [P3] \\
    & Code        & Written pseudo code or code syntax & 23 & def update\_task [P5] \\
    & Annotation  & Symbols and annotations & 31 & circle, arrow, underline \\
    & Freeform    & Sketches or drawings without clear structure & 9  & line chart [P1] \\
\midrule
\multirow[t]{2}{*}{Approach} 
    & One-Time    & Marked all possible changes before generating code edits & 25 & ``add due\_date'' as an attribute and pointing arrow to the written def sort function [P4] \\
    & Step-by-Step & Decomposing tasks and generating code edits after each subtask & 49 & Copied ``list\_task'' function, generated, then edited it to be list tasks by ``due\_date'' [P2] \\
\midrule
\multirow[t]{2}{*}{CodeRef} 
    & Parameter   & Referencing code as a parameter to contextualize generation & 12 & Circled data and pointed to plot [P3] \\
    & Target      & Reference to code as the target to be modified & 28 & Crossed out sampled data with ``(int, int)'' to ``(int, str)'' [P5] \\
\midrule
\multirow[t]{2}{*}{Purpose} 
    & Functional  & Sketches express how the code should function & 11 & Sketching the sample processed out\-put [P6] \\
    & Procedural  & Sketches express how the code should process or run & 63 & Sketching the flow from variables, to one-hot encoding, to distance metrics [P5] \\
\midrule
\multirow[t]{2}{*}{Form} 
    & Concrete    & Sketches with a concrete or syntactic form & 31 & text, pseudo code \\
    & Abstract    & Sketches representing the abstract attributes (semantic) meaning & 43 & annotations, freeform sketches \\
\bottomrule
\end{tabular}
    \caption{Types of sketches used in stage one were categorized, and the number being coded was indicated by $N$.}
    \label{tab:first_code}
\end{table*}


\subsection{Results}
\label{sec:result}

All but two participants completed the four assigned tasks; these two participants did not complete \f{scenario2}-\f{task2} within the assigned time.
There was a concern that experienced programmers might be strongly biased toward typing code edits, which could impact their experience with sketch-based code editing. However, all participants appreciated the concept and expressed a willingness to integrate it into their current programming workflow, as it allowed them to \pquote{think deeper about the code}{P1} and \pquote{focus on higher-level planning}{P6}.
Participants used sketching to edit code an average of 3 times per subtask ($SD=4.0$). \rev{Each instance of sketching often included multiple annotations, with some sketches encompassing edits to several parts of the code.}
Early-stage code edits were primarily made through sketched annotations, but in the later stages (12-13 minutes), edits occurred without sketches, suggesting the use of a keyboard or undo/redo mechanisms to refine code. P2 and P4 explained that they resorted to the native tablet keyboard for edits to handle low-level details, as the waiting time for model interpretation \rev{could exceed five seconds (in average around 4-8 seconds based on the size of the codebase) in some cases, making manual code changes faster}, thus \pquote{would rather do it myself [themselves]}{P4}.

\subsubsection{General Workflow} 
Participants sometimes wrote higher-level instructions first when unsure about the solution but had a rough idea of where the code edits should happen and what the \pquote{shape of the code looks like}{P4}. 
After evaluating the edited code, they then added annotations for lower-level code editing based on their approaches in mind.
We also observed two participants gradually develop a personalized workflow for editing code with sketches. P2 found that breaking down tasks into very low-level details was ineffective and not necessary, while P5 emphasized the need for smaller task pieces for better system understanding. \rev{This difference arose because P2 included precise code-like keywords in their sketches, minimizing the need for further detail.}

\begin{figure*}[th]
    \centering
    \includegraphics[width=\linewidth]{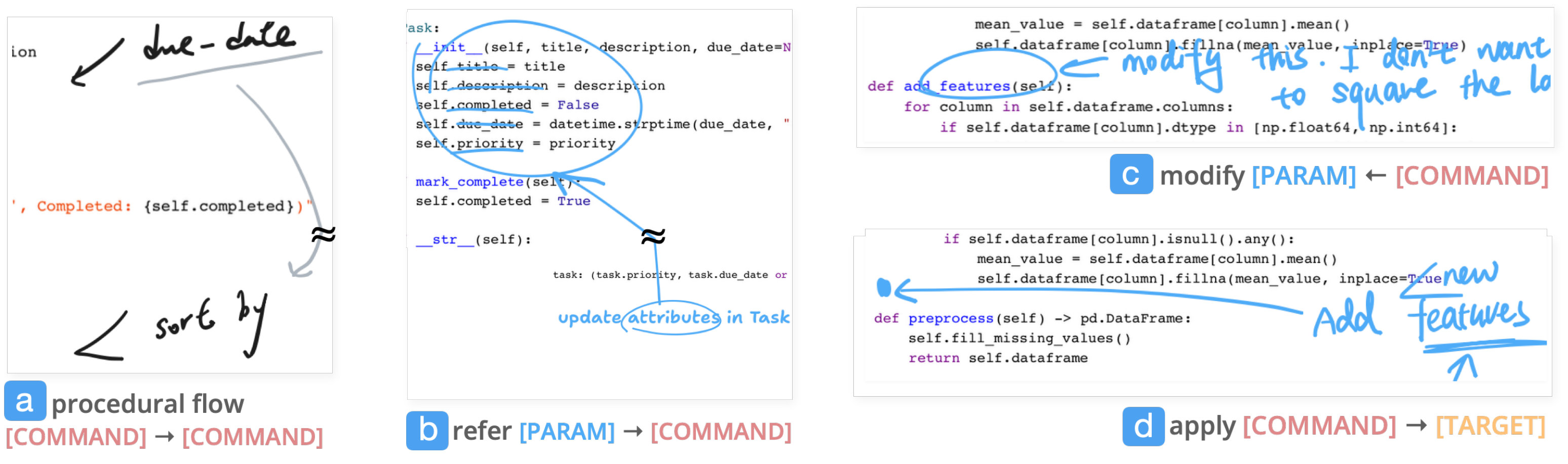}
    \caption{Sketches from stage one showing how participants employ arrows ($\rightarrow$) for different purposes, including command (the intended action of operation), parameter (supplementing the command), and target (the area where the edit should occur); (a) indicating procedural flow between commands; (b) referring to data attributes; (c) modifying a function, with the function as the parameter to supplement the command; (d) applying changes to a target area.}
    \Description[Data collected from the user study showing how participants employ arrows for different purposes]{The figure depicts sketches from a user study, showcasing how participants use arrows to indicate different operations in a code editor. The arrows are employed for three distinct purposes: (1) command, representing the intended action or operation; (2) parameter, supplementing the command; and (3) target, pointing to the area where the edit should occur. The figure is divided into four labeled sections: (a) Procedural Flow: Displays arrows connecting commands, such as "due_date" and "sort_by," to illustrate the procedural flow of tasks; (b) Refer: Shows arrows linking parameters like "attributes in Task" to commands, highlighting how participants refer to specific data attributes; (c) Modify: Demonstrates arrows pointing to a function like add_features, with handwritten annotations such as "modify this" and "I don’t want to square the log," reflecting user-directed modifications; (d) Apply: Includes arrows connecting a command (proprocess) to its target, emphasizing changes being applied to a specific section of the code. This figure captures the interactive coding process, with annotations providing insights into the participants' thought processes and actions during the study.}
    \label{fig:arrow-variants}
\end{figure*}

\subsubsection{Types of Sketches}
Overall, the sketches can be situated in a quadrant with two spectrums (\autoref{fig:classification}): \f{Abstract}-\f{Concrete} and \f{Procedural}-\f{Functional}.
The \f{Abstract}-\f{Concrete} spectrum describes whether the annotations are abstract symbols or graphs versus concrete written text. The \f{Procedural}-\f{Functional} spectrum classifies the target of the annotations, ranging from procedural steps describing how the program should be structured to functional descriptions specifying how the program should work. Participants often combined these aspects, drawing graphs and adding arrows to refer to certain data attributes, specifying both functional and procedural terms.

\subsubsection{Sketch as a Tool} 
Additionally, we observed that participants considered sketches as functional ``tools'' that could be reused~\cite{renom2022exploring}, not just as transient digital ink drawings. All participants expressed that they could use different sketches to achieve the same effect, choosing which sketch to use based on the environment, such as available white space. They also reused their sketches to convey the same effect; for instance, an arrow used to insert a function into a specific line of code was reused by P3 to add another function.


\subsubsection{Ambiguity of Sketches and Model's Transparency}
The primary challenge was the ambiguity of participants' sketches. For example, arrows were used inconsistently, sometimes indicating context [P1] and other times denoting targets of changes [P4] (\autoref{fig:arrow-variants}a-d).
The interpretation of these sketches often relied heavily on surrounding code, leading to occasional misrecognition and misinterpretation. 
This necessitated an iterative refinement process.
However, this iteration became a significant source of frustration for participants, largely due to the system's lack of transparency.
Participants rated the clarity of the effect of their sketches on the generated code poorly ($Mdn=3.5$, $SD=1.83$), as well as the ease of iterating on sketches ($Mdn=4$, $SD=2.34$), \rev{on a seven-point scale questionnaire.}
Participants struggled with not knowing \pquote{where the code was being edited}{P4}, an unclear mapping between sketches and the edited code, and why the model misinterpreted their sketches.
This is considered as interpretation error than recognition error.

\subsection{Summary}
The results revealed that programmers utilized diverse sketching techniques, necessitating an iterative refinement process due to the inherent ambiguity of these sketches.
However, the current iteration process was hindered by the AI model's lack of transparency, particularly in how it interpreted sketches and applied code changes.
To address this issue, we focused on identifying potential misinterpretations of sketches by the AI model and exploring how programmers could recover from these errors in the next stage.

\section{Stage Two: Model Interpretability}
The second stage of our study focused on enhancing user control over the model interpretation of sketches by providing different types of brushes for sketching and adding feedback to convey the model's interpretation of the sketches. 

\subsection{User Interface}
\begin{wrapfigure}{l}{15mm}
\vspace{-3mm} \includegraphics[]{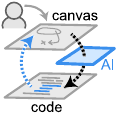}
\end{wrapfigure}
\begin{figure*}
    \centering
    \includegraphics[width=\linewidth]{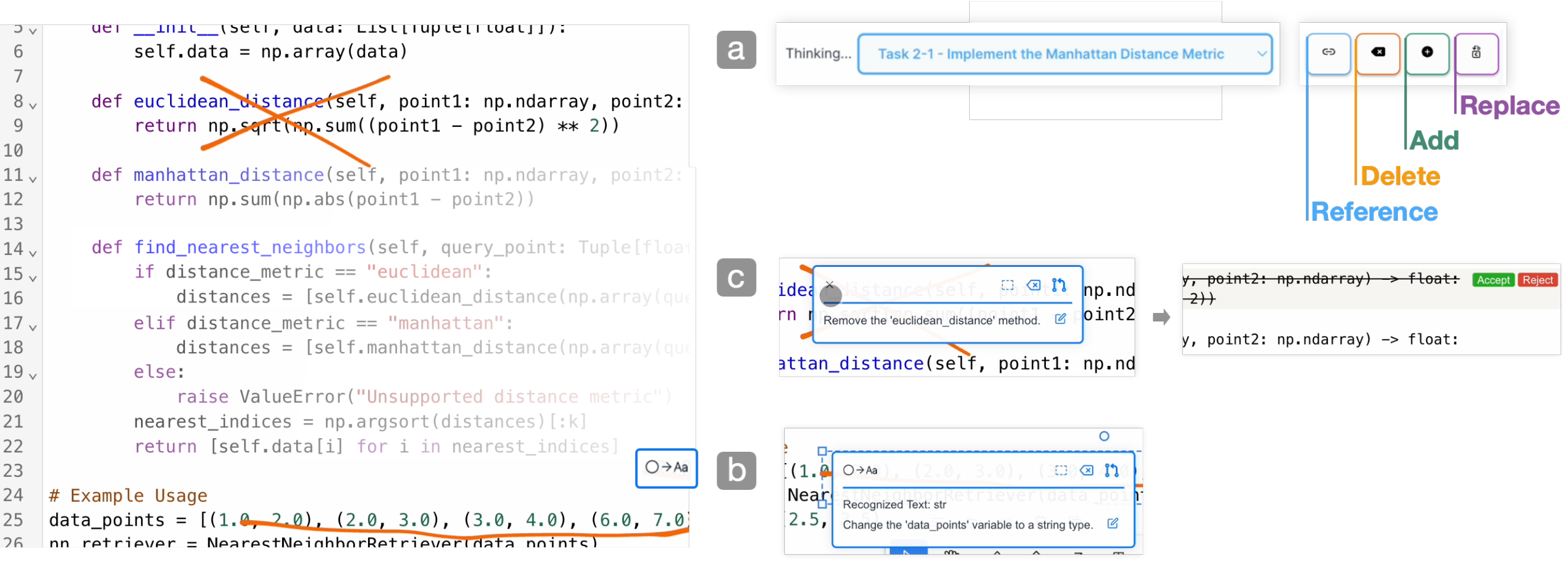}
    \caption{The interface design from the second stage. (a) Command brushes used to steer AI model's interpretation, including reference, delete, add and replace; (b) interpretation of sketches displayed as tooltips, programmers can click on the preview (recognized sketches) and see the full description of AI's reasoning of actions; (c) programmers can click on the commit button to execute the actions, edited code will be shown in diff view and programmers can accept/reject it.}
    \Description[Stage 2 interface with a code editor, AI-assisted features and sketch interpretation.]{The figure showcases the interface design from stage two of the study, illustrating an AI-assisted Python code editor for the NearestNeighborRetriever class. The interface highlights several interactive features: (a) Command Brushes: Located at the top right, labeled "Reference," "Add," "Delete," and "Replace," these tools allow users to guide the AI’s interpretation and actions within the code. (b) Tooltip Display: A tooltip is shown, presenting the AI’s interpretation of a sketched annotation, including an option to view the full reasoning behind the AI's suggested changes. For example, a tooltip highlights a change suggesting converting a variable to a string type. (c) Commit Button and Diff View: Programmers can click the commit button to execute the AI's suggestions. The modified code is displayed in a diff view with options to accept or reject individual changes. The interface also includes line numbers, syntax highlighting, and handwritten annotations such as "str" near the code. A task description at the top prompts users to "Implement the Manhattan Distance Metric," guiding the focus of the current programming task. This design facilitates a collaborative and iterative code development process between the user and the AI system.}
    \label{fig:second-interface}
\end{figure*}

 
The system was enhanced with three major features to facilitate the above focus of our second-stage study. 
First, \textit{command brushes} were introduced to allow programmers to convey their intentions more precisely. For example, a ``replace'' brush can instruct the model to limit its interpretation to replacing existing code with the users' sketches (\autoref{fig:second-interface}a).
Second, the underlying model interpretation mechanism was modified to recognize, group, and interpret sketches. The system groups semantically related sketch marks each time the pen is lifted and provides reasoning for the actions it interprets for each group (\autoref{fig:second-interface}b). These descriptions are displayed as tooltips next to each sketch group, allowing users to edit the descriptions to refine the interpretation or commit to executing the actions.
Third, an inline diff view was added to the code editor, enabling users to visualize code changes as staged edits and choose to accept or reject these changes (\autoref{fig:second-interface}c).

\subsection{Participants, Tasks, and Procedure}
Six new participants were recruited through convenience sampling, with all right-handed, 2 identified as women and 4 as men. All participants had 3-6 years of programming experience in Python and had used ChatGPT or Copilot 6-14 times per week. The same scenarios and tasks were used with the same procedure and data collection. 
We applied inductive thematic analysis to the observational notes, screen recordings, system logs, captured sketches, and interview notes to identify common types of model interpretation errors, the strategies participants used to recover from these errors, and insights related to the model’s interpretation.

\subsection{Results}
Four participants did not complete one of the four assigned tasks from either \f{scenario 2} or \f{scenario 3}. This incompletion was acceptable, as our primary objective was to understand how participants recovered from interpretation errors, rather than task completion itself. We identified a total of 66 error and recovery scenarios, categorized into six distinct error types (\autoref{tab:error-strat}). Participants employed six different repair strategies following three major actions: rejecting/accepting code edits, or taking no action.

\subsubsection{Feedback on Model Interpretation}
Most participants (5/6) appreciated having an interpretation as a preliminary step before code edits. They noted it is necessary when code changes went wrong (5/6), when they were unsure about how the code should be implemented (4/6), and when tasks required decomposition (2/6). However, most of the time, they could rely on the code diff view as it indicates the model's interpretation, especially if their sketches included pseudocode. They expressed that the interpretation feedback should include the recognized items and text within the annotation (6/6), the model's recognition of non-textual annotations and sketches (5/6), suggestions for code edits (3/6), and the linkage between sketches and code edits (2/6).

\subsubsection{Common Errors and Strategies to Repair}
Of all sketches, $23.2\%$ required iterations due to model interpretation errors.
Common errors (\autoref{tab:error-strat}) included mismatches between code implementation and user expectations, incorrect interpretation before code edits, wrong recognition of sketches, no code edits being made, incorrect scope of code changes, and wrong modified code syntax causing runtime errors.
The most frequent error was code mismatches, which were detected after code edits, P10 questioned whether the error happened \pquote{because my drawing was not clear enough or my [written pseudocode] was not recognized.}
The second and third most common were incorrect interpretations of user actions and recognitions, which participants could identify before any code changes occurred. In these cases, some participants (4/6) chose to refine their sketches before generating the code edits, while two participants occasionally still pressed the generate button, P11 explaining this due to \pquote{not knowing how should I refine the sketches.}

In most cases, participants attempted to repair errors by redrawing sketches. In two instances, they edited the code directly using the tablet's keyboard, in three cases they adjusted the interpretation, and in four cases they used control brushes. However, participants only used the control brush when redrawn sketches were still not recognized. P9 explained, \pquote{I would think that it's because of the recognition error or code referencing error, then realize it's misinterpreting what I want to do.}
Overall, these repair strategies can be categorized into three types: selection, instruction, and target. These included adding textual instructions, adding annotations, removing unnecessary sketches, rewriting pseudocode, adding code syntax, and adding references pointing to other target code (\autoref{tab:error-strat}).
For instance, P9 changed the written text from \pquote{handle} to \pquote{def} to specify that the handling should be implemented in a new function. 
Some participants redrew sketches to prevent new annotations from obscuring the code when no sketches were detected, suspecting that \pquote{the sketches blended into the code}{P11}. \rev{This concern is valid, as the system overlays the sketch layer on top of the code layer in the pictorial form to associate sketches with specific lines of code during interpretation.}
All participants used strategies such as adding code targets and references to \pquote{make sure correct code is being used}{P7} or \pquote{only changing specific area [of code]}{P8}.
For instance, P8 circled the DataProcessor class to ensure that new code edits were implemented as methods within the class rather than as standalone functions outside it.

\begin{table*}[hbt]
    \caption{Results from stage two showing AI interpretation error types and corresponding participant repair strategies.}
    \centering
    \small
\begin{tabular}{
        >{\raggedright\arraybackslash}m{1.5cm}
        >{\raggedright\arraybackslash}m{3.2cm}
        >{\raggedright\arraybackslash}m{.5cm}
        >{\raggedright\arraybackslash}m{1.2cm}
        >{\raggedright\arraybackslash}m{1.5cm}
        >{\raggedright\arraybackslash}m{2.9cm}
        >{\raggedright\arraybackslash}m{.5cm}
        >{\raggedright\arraybackslash}m{3cm}
    }
    \toprule
    \multicolumn{4}{l}{\textbf{ERRORS}} & \multicolumn{4}{l}{\textbf{REPAIR STRATEGIES}} \\ 
    \midrule
    \textbf{Type} & \textbf{Description} & \textbf{$N$} & \textbf{Action} & \textbf{Type} & \textbf{Description} & \textbf{$N$} & \textbf{Example} \\ 
    \midrule
    Code Mismatch & code does not match intended implementation & 25 & Reject & Add Code Target & specify target of code changes & 15 & \includegraphics[height=45pt]{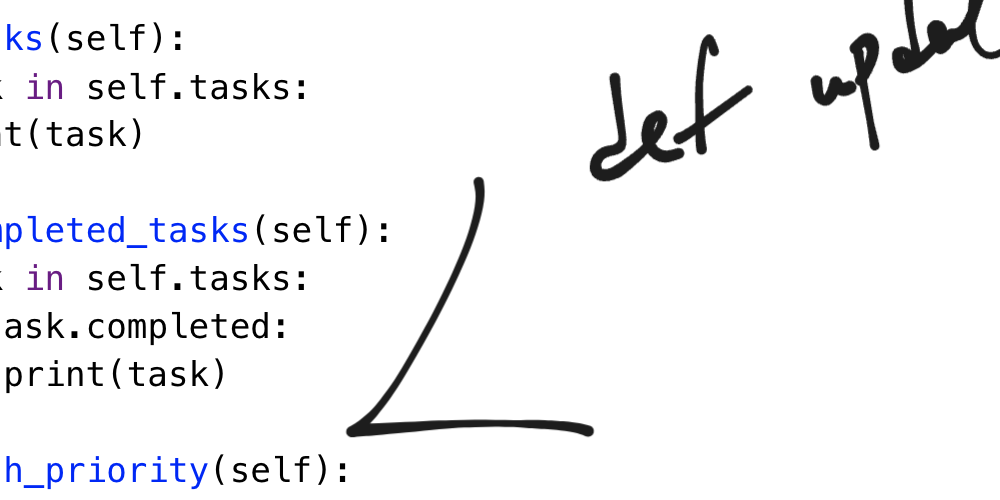} \\ 
    Code Error & contains syntax or logical error & 3 & & Add Code Reference & add context for generation & 11 & \includegraphics[height=45pt]{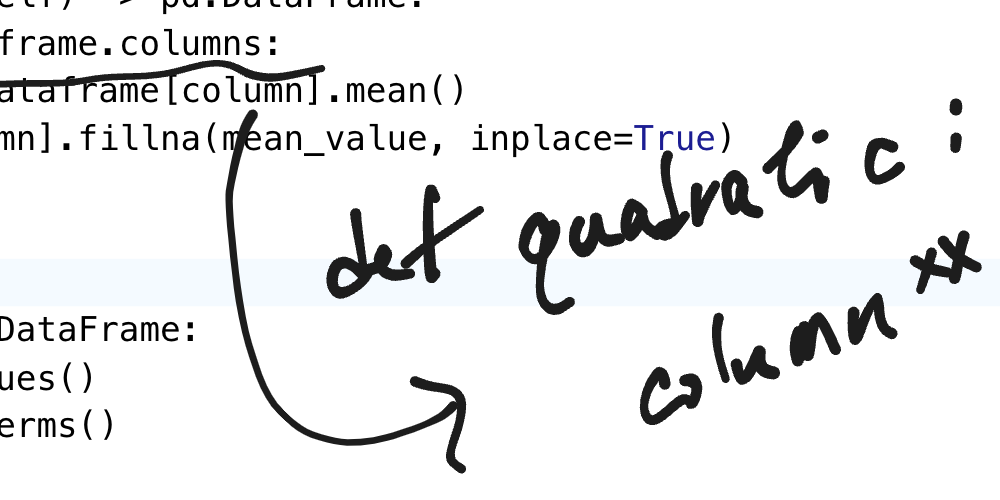} \\ 
    No Changes & no code changes being made & 7 & & Precise Annotation & rewrite text or redraw annotations & 11 & \includegraphics[height=45pt]{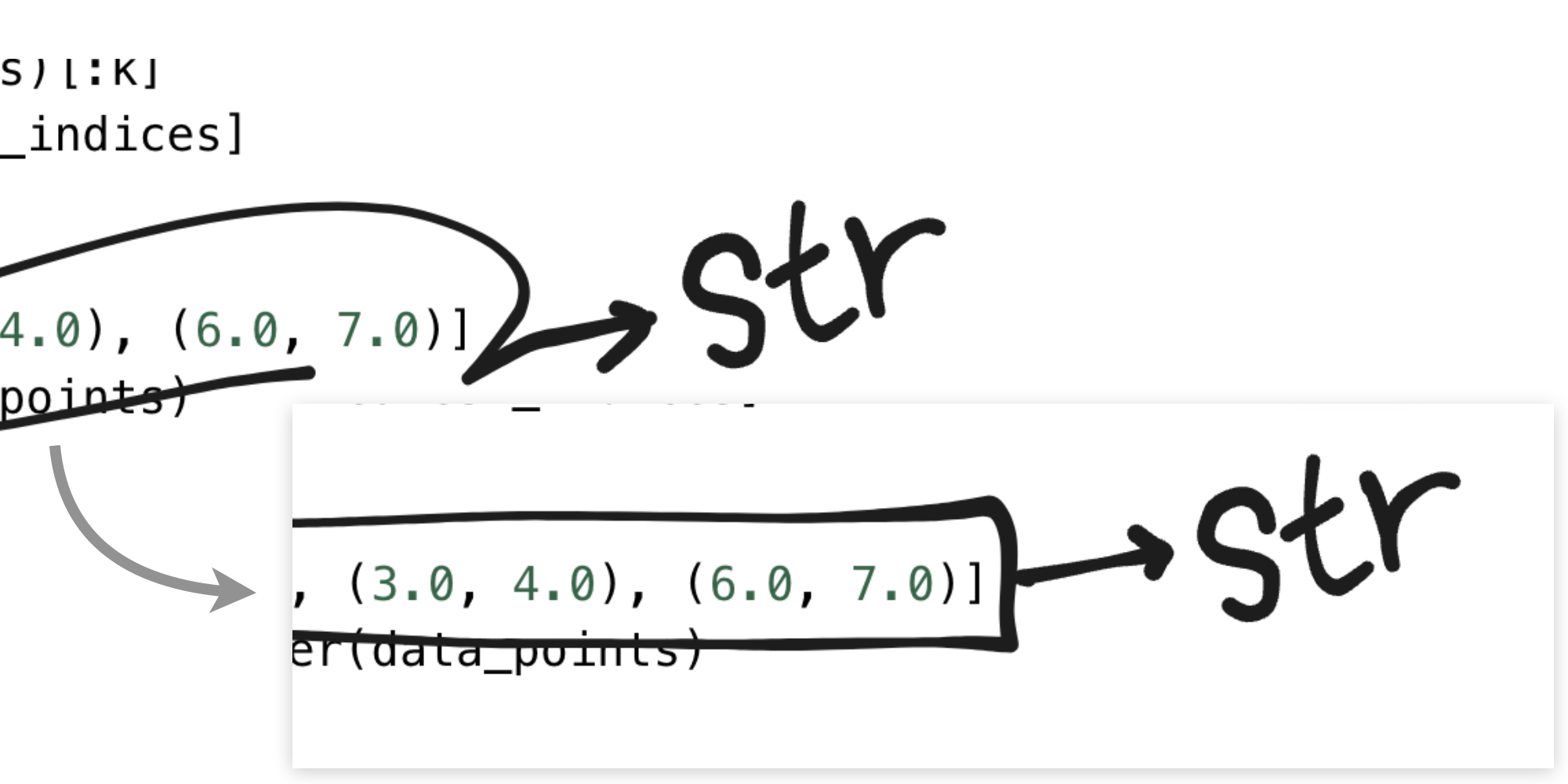} \\ 
    Wrong Action Interpretation & interpret the action wrong before generate code edits & 14 & & Add Pseudo Code & add code syntax or symbols & 19 & \includegraphics[height=45pt]{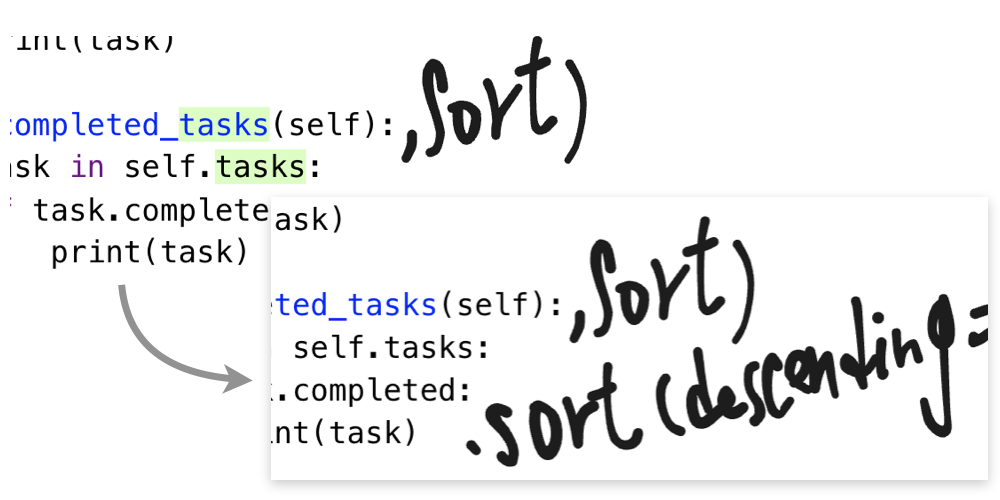} \\ 
    \cline{1-4}\cline{5-8}
    Wrong Scope of Change & incorrect code range edited & 4 & Accept & Next Round & accept code edits then annotate again & 4 & \\ 
    \cline{1-4}\cline{5-8}
    Wrong Recognition & recognize the sketches wrong & 13 & No Action & Regenerate & generate code again without modifying sketch & 6 &  \\ 
    \bottomrule
\end{tabular}
    \label{tab:error-strat}
\end{table*}

\subsubsection{Control Over Model Interpretation}
Most participants (5/6) did not find control brushes particularly useful. Two participants preferred interacting directly with the code editor rather than using specific brushes to constrain the AI model. They favoured simple arrows and cross-outs to indicate code replacements instead of different brushes. All participants found that sketches alone were expressive enough to guide the AI model in correcting its interpretation.
For example, P10 added a numbered label to the pseudocode, \pquote{$\rightarrow$ str:}, to indicate that the AI should prioritize interpreting that annotation first \includegraphics[height=14pt]{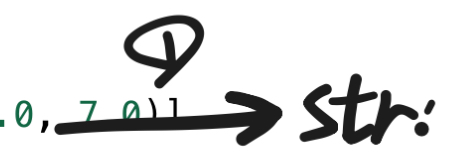}.
We observed that three participants tended to wait for the interpretation to complete before generating code edits to \pquote{not lose control over my [their] own code}{P9}.

\subsubsection{Sketching, Correcting Model, and Editing Code}
All participants primarily attributed their frustration with iteration to the need for context \pquote{switching between the code editor and the canvas}{P9}. The interface required a double-tap to enter or exit the code editor for tasks such as accepting/rejecting code edits, undoing/redoing actions, and performing manual edits (though these were less common). Due to the dynamic nature of programming, where each edit builds on previous modifications, the frequent need for interpretation and the requirement to accept or reject code edits often disrupt participants' flow. Consequently, most participants (5/6) preferred to complete all sketches first and then use the ``generate'' button as a clear boundary between debugging and sketching modes, avoiding repetitive context switching.

This context switching also involved changing mental models and using different input modalities, leading to errors. For example, some participants (3/6) frequently selected the wrong tools due to overlapping semantic meanings, such as P11 using the eraser to delete code or the pointer to select code. These findings highlight the importance of enabling interactions with the code editor ``through'' the canvas layer, effectively translating certain canvas layer interactions into actions within the code editor layer.


\subsection{Summary}
While feedback on AI interpretations added value, the method used in this stage disrupted the programmers' flow. The goal of code shaping is to allow programmers to edit code structure through sketches, rather than engage in low-level code editing or prompt engineering for the AI system. The control brushes did not perform as expected; participants preferred refining their sketches by adding more code references or employing code syntax to shape the outcome. This tendency can be linked to the cognitive dimension of \emph{premature commitment}---forcing programmers to make decisions too early~\cite{green1996usability}, which conflicts with the iterative nature of code shaping. The findings underscore that the key to facilitating code shaping is an interaction design that minimizes the conceptual layers between the code editor and the sketching canvas.

\section{Stage Three: Towards Reconciling Sketches, AI, and Code}
To bridge the conceptual gap between interacting with the code editor and the canvas, the user interface was modified in two key ways. Insights from the previous stage suggested that unnecessary code changes could be minimized by confirming interpretations before generating edits. Building on the types of interpretations identified, we introduced an always-on feedforward mechanism through subtle visual cues. This approach allows programmers to iterate more quickly without delving into code details.
Additionally, to reduce the cognitive load of switching between layers, we developed unique gestures that enable users to interact directly with the code editor through the canvas.

\subsection{User Interface}
\begin{wrapfigure}{L}{15mm}
\vspace{-3mm} \includegraphics[]{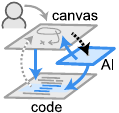}
\end{wrapfigure} 
Unnecessary GUI elements were removed to keep the interaction focused on sketches. The button for sending sketches and code to the model was renamed ``Commit'' to make it clear that a change will be applied to the code. Only this button and the ``Run'' button were retained in the GUI, as participants preferred having explicit controls for these actions rather than relying on implicit gestures. We open sourced the code for this stage at \url{https://github.com/CodeShaping/code-shaping}, including all the prompts used.

\begin{figure*}[t]
    \centering
    \includegraphics[width=\linewidth]{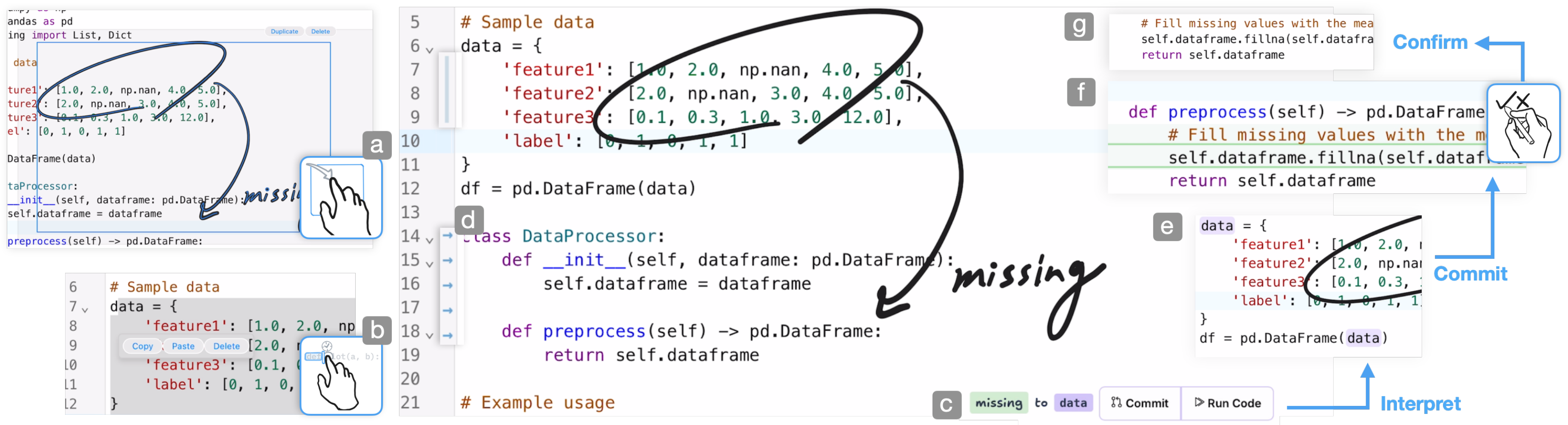}
    \caption{The interface design from the third stage. (a) programmers can use one finger tap and drag to select items on canvas; (b) tap longer and drag will select code with contextual menu beside; (c) the always-on feedforward interpretation showing the interpretation of sketches or text, the reasoning of action, and the related code; (d) the gutter will be decorated to indicate which code being referenced and which code will be affected; (e) the related code syntax will be highlight transiently; (f) programmers can commit the changes and (g) draw check/cross to accept/reject code edits.}
    \Description[Interface design for stage three of AI-assisted code editing.]{The figure illustrates the interface design from stage three of the study, showcasing AI-assisted code editing for a Python data processing task. Key interactive features are labeled (a) through (f): (a) Finger Gesture for Selection: Users can tap and drag with a finger to select items on the canvas, enabling precise interaction with the code; (b) Contextual Menu: When code is selected via a long tap and drag, a contextual menu appears, providing options such as "Copy," "Paste," "Delete," and more; (c) Always-On Feedforward Panel: An interpretation panel continuously displays the AI's reasoning, including sketches or text annotations and the corresponding proposed actions; (d) Decorated Gutter: The code gutter is highlighted to indicate which sections of code are referenced or affected by the proposed changes; (e) Transient Syntax Highlighting: Relevant code is temporarily highlighted to visually connect AI suggestions with specific parts of the code; (f) Commit Button: A commit button allows users to finalize edits. Users can accept or reject changes using checkmark or cross gestures directly on the interface; (g) The interface displays Python code for a DataProcessor class, including a sample dataset and a preprocess method that fills missing values in the dataset. Handwritten annotations, such as the word "missing," and gestures, like arrows connecting actions, are visible throughout, illustrating the interactive coding process.}
    \label{fig:third-interface}
\end{figure*}

\subsubsection{Ink and Gestures}
Based on insights from the previous two stages, we classified common interactions during code shaping into five key categories: navigating the canvas and code, undoing and redoing actions such as code edits or sketches, selecting code or sketches on the canvas, accepting and rejecting code changes, and creating free-form sketches (\autoref{tab:gestures}).
Multi-touch gestures were assigned to system-level interactions, such as panning for navigation and two- or three-finger double-tapping for undo and redo actions. Selecting items on the canvas or within the code editor was differentiated by the duration of a single touch: a single tap for canvas items selection (\autoref{fig:third-interface}a) or a long press followed by dragging for code selection (\autoref{fig:third-interface}b). Contextual action buttons, such as delete and copy, are displayed next to the selected code or within the selection box of canvas items, allowing for quick access to common actions.
We implemented unique stroke gestures for accepting or rejecting code edits using the \$1 unistroke recognizer~\cite{10.1145/1294211.1294238} to detect check (\faCheck) and cross (\faTimes) marks (\autoref{fig:third-interface}f). The Google Cloud Vision API was employed for robust recognition of handwritten text, enhancing the system's ability to interpret written pseudo code or textual instructions.

\subsubsection{Always-On Feedforward Interpretation}
\label{sec:feedforward}
Building on insights from the second iteration, we focused on providing only three essential types of interpretations that users truly needed: (1) recognizing how the model interpreted written text, code, and annotations (\autoref{fig:third-interface}c); (2) describing the code editing action inferred by the model; and (3) indicating the code context by highlighting relevant parameter code, displaying blue vertical line glyph decorations, and marking potentially affected code areas with a $\rightarrow$ icon beside the line number on the glyph (\autoref{fig:third-interface}d).
To identify the relevant code, we traversed the abstract syntax tree (AST) to dynamically highlight code syntax related to the user's input. The interpretation process is triggered 500 milliseconds after the user stops sketching, ensuring timely feedback while minimizing disruptions (\autoref{fig:third-interface}e).
The average latency between the input request and the complete output measured from the second study is approximately $2.87$ seconds (\(SD = 1.45\)). However, since interpretations are generated in real-time as users are sketching, it is possible for the system to produce correct results even before all annotations are fully completed.
We implemented a cascade interpretation approach, sequentially processing pen or touch input, predefined gestures, text and shape recognition, code edit action reasoning, and affected code analysis. This approach enables programmers to adjust their sketches concurrent with system evaluation, rather than waiting for the final step. 
These feedforward interpretations were not directly displayed on the canvas or situated within the code but were instead ambiently presented, updating on the fly to offer guidance when needed.
To enhance usability, especially for right-handed users, we repositioned the interpretation text from the upper right to the lower right of the screen. This adjustment allows users to occlude the interpretation with their hands while sketching, minimizing interference with their workflow.

\subsection{Procedure and Data Analysis}
We recruited six new participants for this study, including five right-handed individuals, four of whom identified as women and two as men. They had between 2-8 years of programming experience in Python and used ChatGPT or Copilot 4-12 times per week. We reused the same setup and tasks to ensure consistency in our results across studies. We added several system logs for gesture recognition and recorded input images, which served as parameters for the always-on feedforward interpretation. We collected $187$ sketches, $48$ of which were recorded when participants hit the ``Commit'' button.
We employed inductive thematic analysis to examine all collected data, including sketches, system logs, video recordings, observational notes, and transcribed interview audio recordings. The iterations were defined by the accomplishment of subtasks that participants themselves decided upon and decomposed from the main study task's goal. We then categorized the common flow of actions within these iterations.

\begin{table*}[bt]
    \caption{The assigned touch and pen gestures to the third stage, enabling interaction across both code editor and canvas layers.}

\small
\begin{tabular}{>{\centering\arraybackslash}m{2.15cm} >{\centering\arraybackslash}m{2.15cm} >{\centering\arraybackslash}m{2.15cm} >{\centering\arraybackslash}m{2.15cm} >{\centering\arraybackslash}m{2.15cm}| >{\centering\arraybackslash}m{2.15cm} >{\centering\arraybackslash}m{2.15cm}}
\toprule
\multicolumn{5}{c}{\textbf{Touch}} & \multicolumn{2}{c}{\textbf{Pen}} \\ \midrule
Navigation & Undo command & Redo command & Select code & Select canvas objects & Accept / Reject code edits & Free-form sketching \\ 
\midrule
Two fingers pan & Two-finger double tap & Three-finger double tap & One finger long press then drag & One finger drag & Check (\faCheck) / Cross (\faTimes) & drawing \\ 
\midrule
\includegraphics[height=47pt]{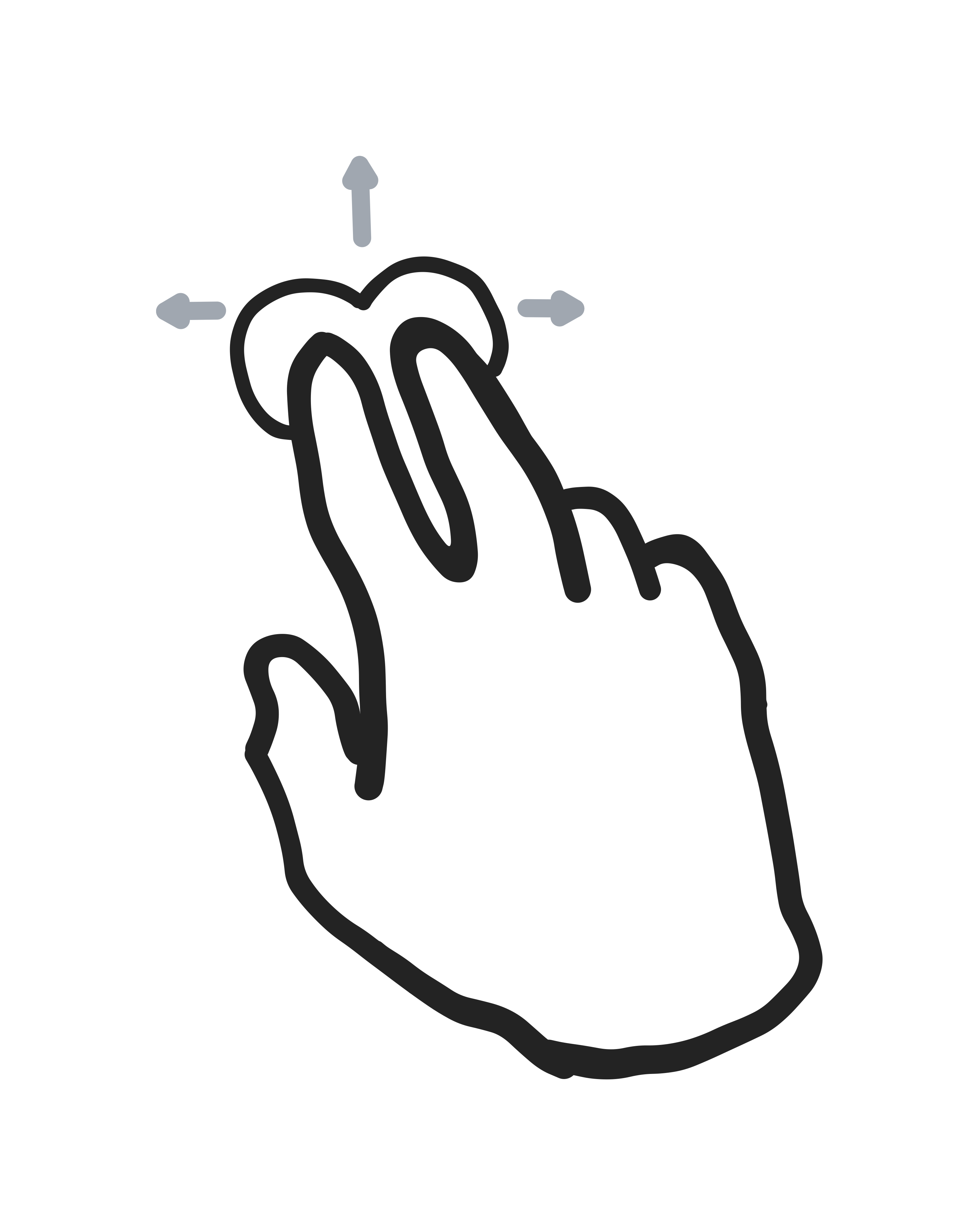} & 
\includegraphics[height=37pt]{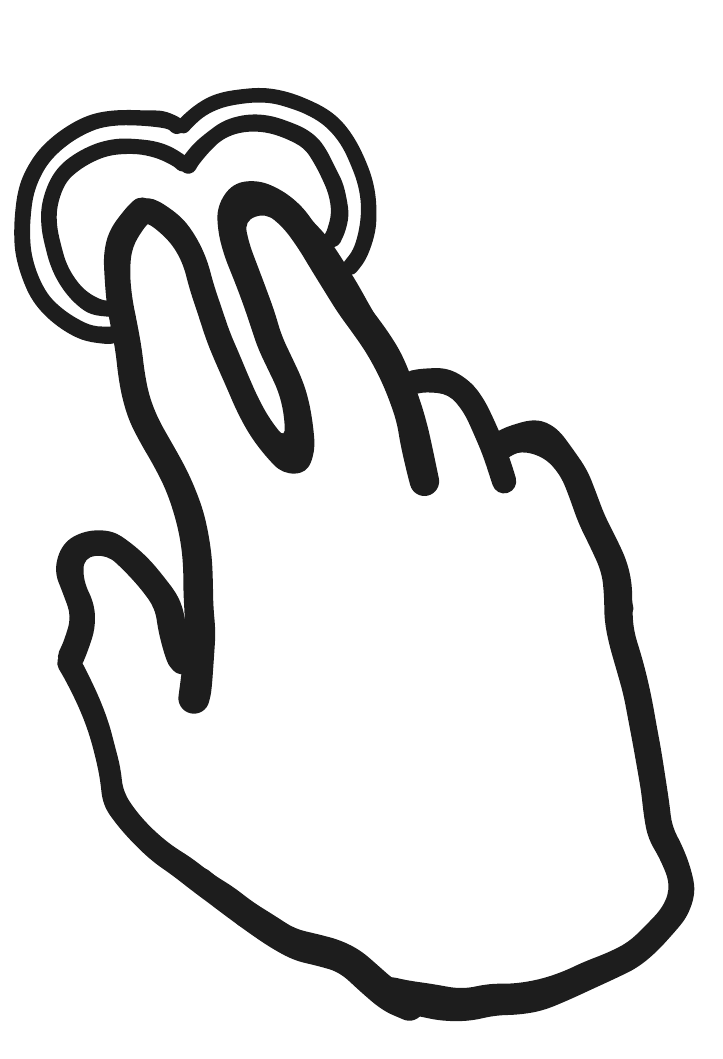} & 
\includegraphics[height=35pt]{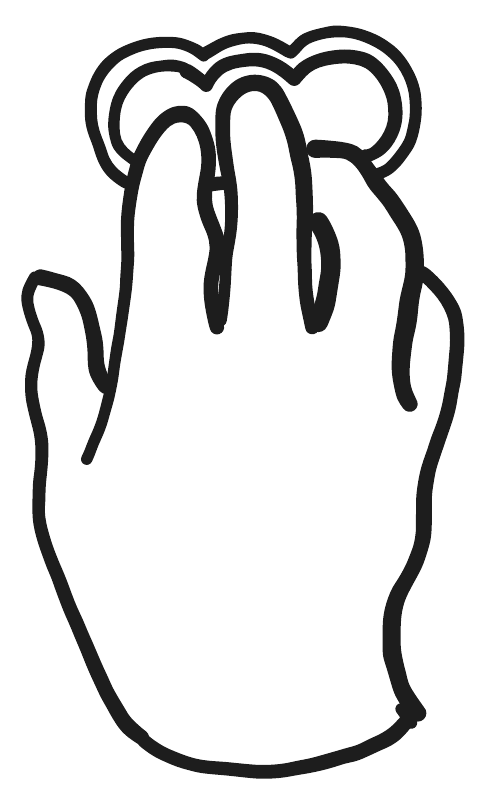} & 
\includegraphics[height=45pt]{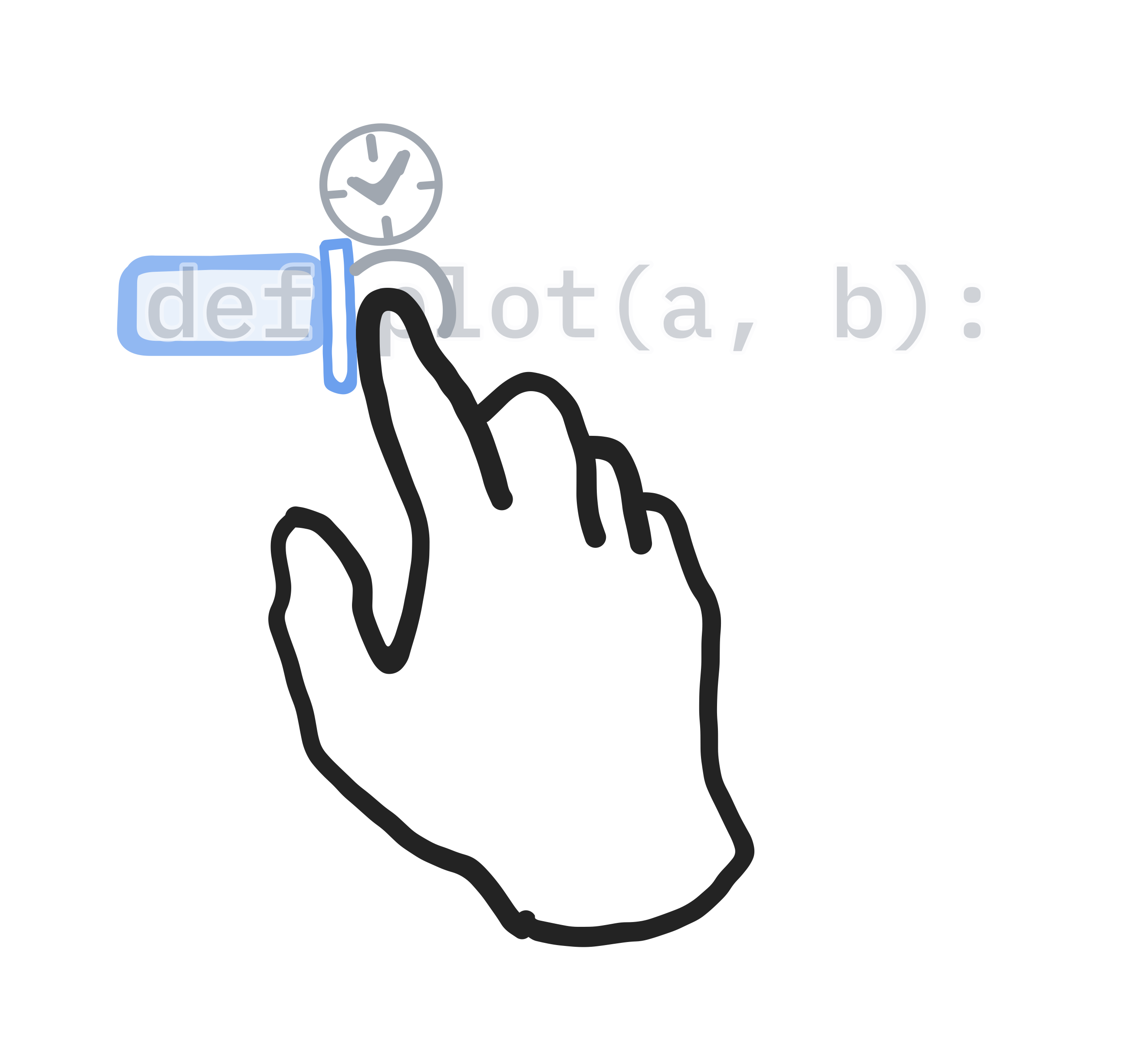} & 
\includegraphics[height=45pt]{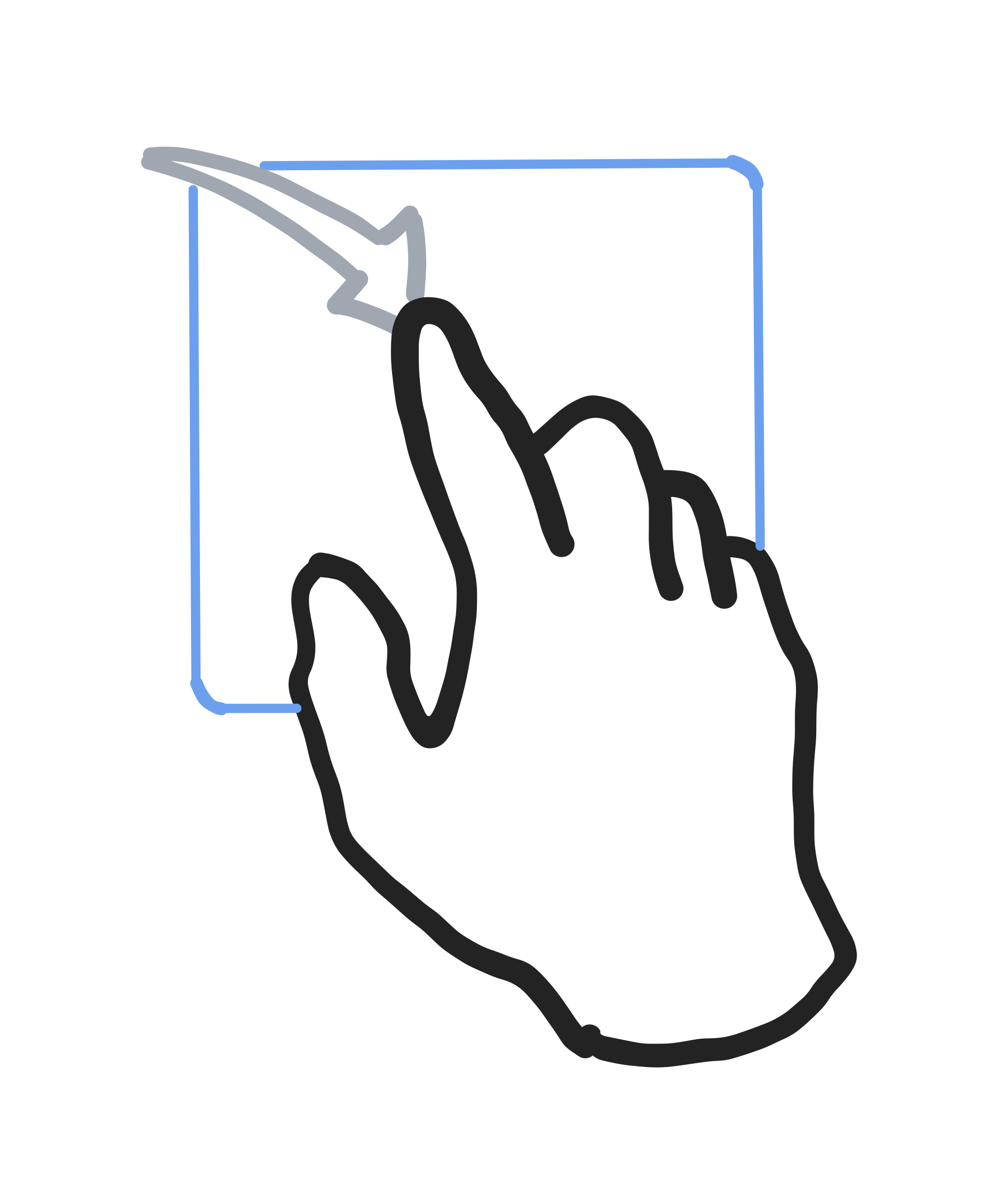} &   
\includegraphics[height=42pt]{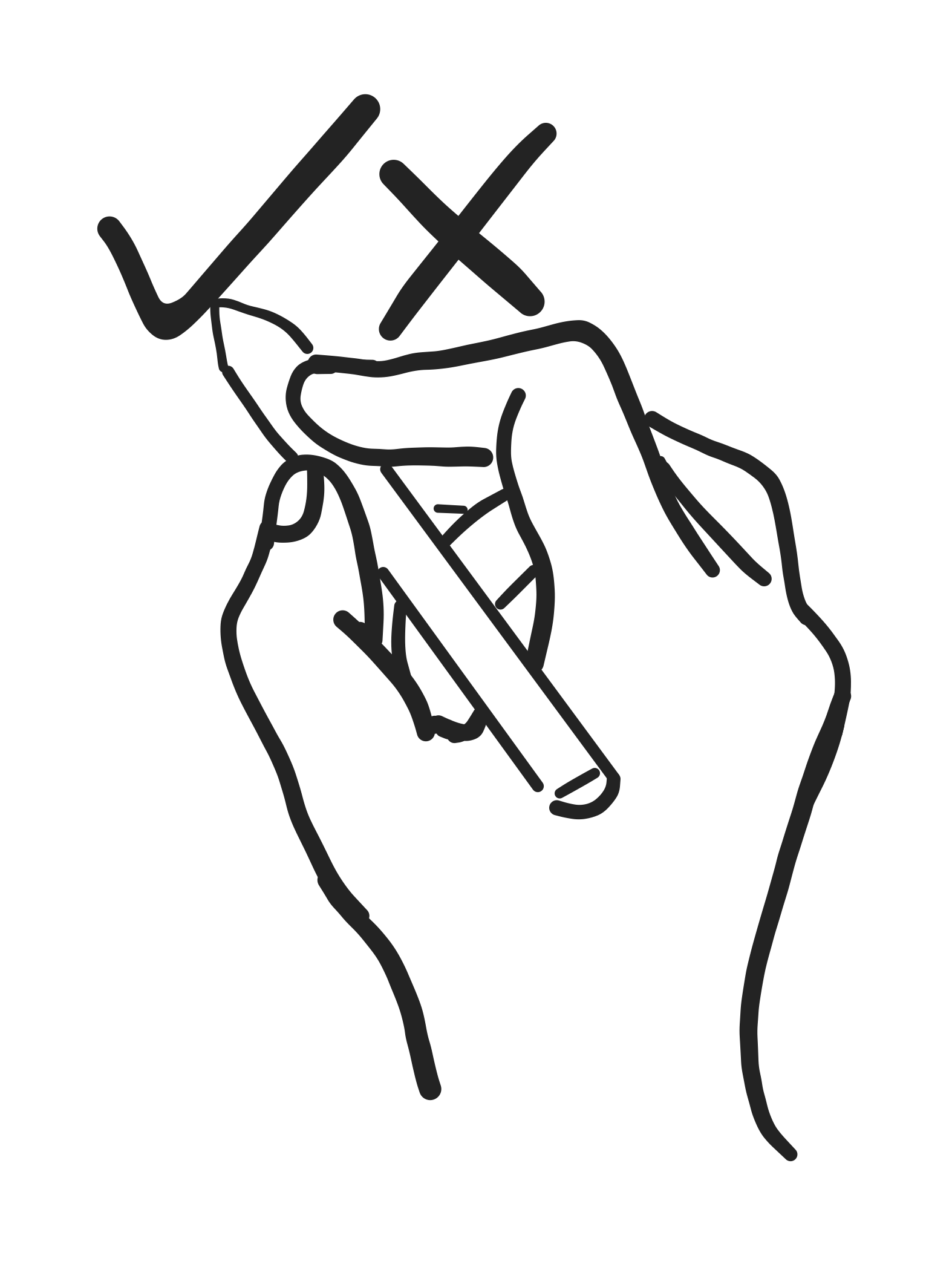} & 
\includegraphics[height=42pt]{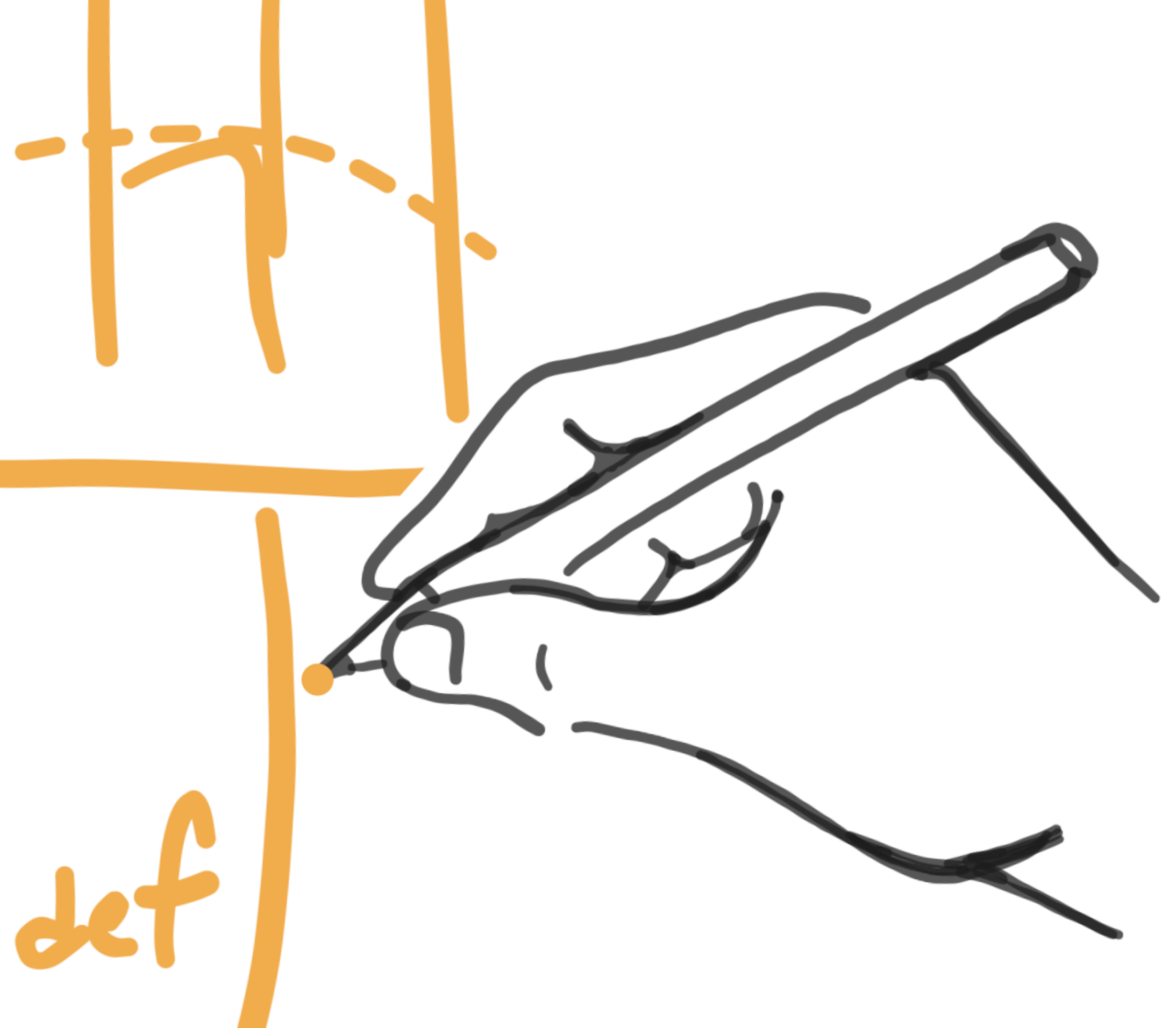} \\ 
\bottomrule
\end{tabular}
    \label{tab:gestures}
\end{table*}

\subsection{Results}
The analysis revealed several themes that shaped participant experiences.

\subsubsection{Flow of Actions}
We identified seven common action flows among participants, highlighting patterns in how they navigated between sketching, code editing, and reviewing interpretations (\autoref{tab:flow}). These flows generally followed a sequence that can be counted as a full iteration:
\[
\begin{array}{c}
\text{Sketch} \rightarrow (\text{Interpret}) \rightarrow \text{Generate} \rightarrow (\text{Run Code}) \\
\rightarrow \text{Accept/Reject}
\rightarrow \text{Re-Sketch/Undo/Redo}
\end{array}
\]

Some of these flows were also observed in the previous stages but were more pronounced in this stage. Participants appeared more aware of their workflows, especially during interviews when recalling their processes. This contrasts with earlier iterations, where participants often expressed uncertainty, such as \pquote{hopefully the code edits are correct}{P8}. They chose different methods to iterate when code edits were incorrect, adapting their actions based on the situation. For example, P14 mentioned using the undo/redo function (ID 4 in \autoref{tab:flow}) for interpretation errors, while opting for the re-sketch approach (ID 3 in \autoref{tab:flow}) in other cases.

\subsubsection{Always-On Feedforward Interpretation}
After viewing the interpretation, participants pressed the ``Commit'' button $32.4\%$ more frequently in the third stage compared to that in the second. P15 explained, \pquote{the interpretation shows what code is possibly being affected is useful to make sure it will not mess up my code}. This underscores the value of glyph decorations in the always-on interpretation, as they may help participants confirm their intended code changes. Three participants echoed sentiments from the second iteration, appreciating the control provided by the interpretation. P16 noted, \pquote{I feel more confident that it’s on the right track [...] I don’t want it to be a black box}. 

Four participants stressed the importance of waiting for the correct interpretation before committing to changes, even if it required slightly more time compared to directly pressing the commit button. P13 explained, \pquote{I would rather wait a bit longer than evaluate the wrong generated code edits}, reflecting a preference for clarity over speed. 
The feedforward interpretation also guided participants on their next steps, regardless of whether the interpretations were correct. For instance, P15 noted that a previously correct target code section became incorrect after adding an arrow for code reference, indicating a misinterpretation of the arrow.

\subsubsection{Sketching or Editing Code}
A key goal of this design was to reconcile the conceptual layers between the code editor and the canvas.
While all participants did not report difficulty switching between contexts, they perceived them as distinct. As P14 observed, \pquote{I still think of the code and annotations separately in my mind}. However, participants found the unique gestures and strokes for interacting with the code editor \pquote{straightforward}{P17} and felt that it \pquote{makes me [them] feel like the sketch is affecting the code}{P18}.

A notable improvement was that most participants (5/6) expressed no need to use the keyboard, even for simple edits like deleting a line of code. When asked why they preferred using a strikethrough to indicate deletion instead of using the backspace key, P14 explained, \pquote{I just want to use sketches and annotations as the only way to change my code}. This suggests a sense of embodied interaction, something reinforced by P12, \pquote{It’s like my hands are directly editing the code based on how I want the code to be.} However, the predefined \faTimes ~gesture for rejecting code edits was triggered by accident once when P17 wrote \pquote{x} as part of the sketch.

\subsubsection{Conceptual Shift in Code Editing}
\label{sec:conceptual_shift}
Participants demonstrated a shift from linear to spatial thinking in their code editing processes. As P16 observed, \pquote{I'm no longer just writing line by line [...] I'm arranging my thoughts spatially}. This reflects a move away from a traditional, syntax-focused approach to one that emphasizes the overall structure and flow of the code. Another participant, P14, highlighted this shift, stating, \pquote{it’s more about the higher-level structure and flow of the code as a whole}.

\subsubsection{Freedom and Flexibility}
The iterative process enabled by free-form sketching provided participants with a sense of creative freedom. P15 reflected, \pquote{This lets me play around with ideas in a way that’s more fluid and creative [...] I'm experimenting}. All participants mentioned that they would often resketch the code edit even when the generated edits were correct, as they discovered better ways to tackle the task. P14 noted that the canvas and undo/redo mechanisms allowed them to \pquote{draw whatever we [they] want and see how the code changes}. 

However, while participants valued the freedom of sketching, they also tended to ``compromise'' based on the AI’s interpretation capabilities. For instance, P18 consistently used squares instead of circles because \pquote{the rectangle works better} and did not obscure the code. As a result, participants developed preferences for specific annotations with the system along the time. P14, for example, switched to using crosses after observing that strikethroughs occasionally applied to the wrong lines of code. Over time, these preferred annotations became interchangeable in practice, as participants felt they \pquote{expressed the same meaning}.

\subsection{Summary}
The third stage highlights the effectiveness of defined gestures and always-on feedforward interpretation in reducing the transmission gap between the conceptual layers of the canvas, code, and AI model. This design iteration demonstrated how to display the model's interpretation, and designed interactions that minimize the need for layer switching.
While minor challenges remain, such as the rare misinterpretation of sketched gestures and some persistent between AI interpretation and actual applied code edits, these issues can likely be addressed by advancements in AI models.

\section{Example Use Case Scenarios}
We demonstrate how code-shaping could integrate with current programming practice in two usage scenarios. The interactions and user interface features to support these are real, we only had to modify the study prototype to support multiple files. Back-end infrastructure, like syncing code with a desktop editor is not implemented. Also see the accompanying video to view these scenarios.

\subsection{Programming on the Couch}
\rev{ Alicia, a data scientist, is improving a machine-learning preprocessing pipeline distributed across multiple files. She wants to introduce flexible scaling and proper categorical encoding for both training and testing datasets. Seeking a fresh perspective, Alicia grabs her tablet and opens the Code Shaping editor to explore solutions.

To start, Alicia opens the editor support code shaping paradigm and runs the current code. She observes that the output does not handle categorical data correctly. Beside the data processing pipeline code, Alicia draws a flowchart directly on the tablet’s screen, visually aligning sketches to the vertical layout of the code. This flowchart proposes a branching structure starting from the code, \inlinecode{def preprocess\_pipeline} $\rightarrow$ 
\includegraphics[height=14pt]{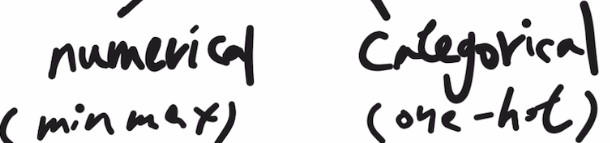}
$\rightarrow$ \inlinecode{processed data}.
The system’s feedforward interpretation (\autoref{fig:third-interface}c) \includegraphics[height=12pt]{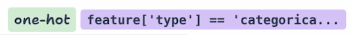} and the gutter (\autoref{fig:third-interface}d) highlights the affected lines, showing that the code will now include a one-hot encoding step where previously categorical features were ignored \includegraphics[height=10pt]{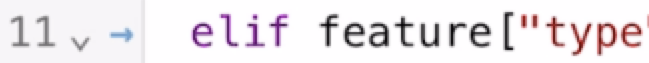}. Alicia commits these changes using the commit button (\autoref{fig:third-interface}f) and then re-runs the code. The updated pipeline applies one-hot encoding to categorical features as intended. However, Alicia notices the code still ignores the \inlinecode{scaling} parameter.

To address this, Alicia decides to refine her sketches without losing her previous changes. She uses a one-finger tap-and-drag gesture (\autoref{fig:third-interface}a) to select the existing flowchart elements. She taps and drags downward on part of the numerical branch to create space and adds a new annotation: \emph{min-max $\rightarrow$} \includegraphics[height=10pt]{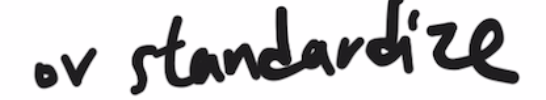}.
The feedforward interpretation and gutter again indicates what text being recognized and which code lines will be altered. Alicia commits these changes, and the editor transiently highlights the updated code snippet (\autoref{fig:third-interface}e). 
With the pipeline now meeting her requirements, Alicia draws a check mark to finalize the changes and remove any temporary sketches (\autoref{fig:third-interface}f). She re-runs the pipeline and confirm that the changes consider the \inlinecode{scaling} parameter. 
}

\subsection{Collaborative Code Reviewing Meeting}
Blair and Carol, senior software developers at a fintech startup, stand before a large interactive whiteboard running the code shaping interface. They are reviewing the core \inlinecode{process\_transaction} function of their payment system.
Blair loads the function into the editor on the board. Carol, stylus in hand, circles a block of nested if-statements for transaction validation. \pquote{These lines are slowing us down} she says, sketching a flowchart beside the code illustrating a streamlined process with early returns.
The system highlights the affected code sections, showing how Carol's sketch translates to code changes. Blair adds to the sketch, drawing parallel arrows for certain validation steps, suggesting \pquote{What if we run these checks concurrently?}.

Carol taps the commit button to see the final edits, but then spots a potential race condition in the new parallel structure. Blair undoes the modifications with a two-finger doubletap, and Carol sketches a new flow with the word \pquote{async} for concurrent validation.
While editing this section of code, Blair notices an opportunity to optimize database queries. He circles lines of code making multiple separate queries and draws a diagram of a batched query approach, consisting of a few boxes representing queries connected by arrows flowing into a single ``batched query'' box.
The AI model modifies the code to use a query builder pattern.
Carol points out that this change might affect error handling. She sketches a new try-catch block structure around the batched query execution. The system modifies the code based on her sketch, with the changes highlighted in green as staged edits.
As they near the end of their session, Blair and Carol review their changes holistically. They use check-marks to accept desired modifications and crosses to reject others, iterating through the highlighted sections.


\section{Discussion}
We discussed participants' multi-level abstraction approaches to shape code, the use of sketches to constrain generated code edits and the design implications of code shaping as a new input paradigm.

\subsection{Interacting with Code Across Multiple Levels of Abstraction}
A program is an inherently abstract entity, lacking a fixed form, and can be conceptualized in various ways—from its tangible output, such as a web page, to the underlying code syntax~\cite{hartmanis1994turing}. 
In this paper, we explored the use of sketches as a medium for programmers to express how they envision code modifications across different levels of abstraction~\cite{10.1145/3526113.3545617, 9127262, 10.1145/3654777.3676357}.
Our initial findings revealed that participants used diverse methods to convey their intentions: some drew visualizations, others provided natural language instructions, and some simply wrote pseudocode. This flexibility stands in contrast to prior methods that rely on one-to-one mappings, such as natural language directly translated to code, predefined visual programming objects, or output-directed programming, where manipulation of output changes the underlying code.

While this paper does not focus on the detailed translation of sketches from various abstraction levels into code, our classification of elements that programmers include in their sketches offers a compelling starting point. 
For instance, in the third stage of the study, we observed that two participants expected the generated code to retain specific function names with underscores as a convention used in Python. However, the AI modified these names to follow a ``camelCase'' format for consistency with other generated code edits. This suggests that while code shaping provides high-level constraints on where and how code edits should occur, the finer details of translating between different abstractions, such as which stylistic elements to preserve, deserve further exploration. Investigating which aspects of sketches should remain consistent and which can adapt in terms of structure or format presents an intriguing direction for future research~\cite{bff9b250-7640-39e2-8f34-329fd1552822}.

\rev{
\subsection{Scope of Sketches}
Code shaping represents the concept of sketching on and around code to perform edits by bridging freeform sketching, AI interpretation, and code. Based on participant tasks, sketches were categorized into commands (intended actions), parameters (supplementary details), and targets (specific areas to edit), see \autoref{fig:arrow-variants}. These sketches often included text, annotation primitives, code syntax, or symbols, and participants occasionally extended them to diagrams, visualizations, or symbolic visuals.

Our findings highlight several tradeoffs in using different sketch representations. First, there is a cost of structure. Participants often preferred minimal-effort annotations that were effective, as creating detailed sketches required significant effort, consistent with information sensemaking \cite{russell_cost_1993}. Second, while the current model can handle many low-level operations (e.g., deleting code, renaming variables, or wrapping lines in functions), participants sometimes opted to type directly for efficiency in Study 1 (e.g., P2 and P4). This suggests a need for integrating primitive gestures, as demonstrated in our third iteration and explored in prior studies~\cite{samuelsson_towards_2023}. Lastly, abstract sketches, though semantically rich, are often difficult to be evaluated and required iterative refinement to align with linguistic code forms. Future research can focus on exploring other types of feedforward interpretation introduced in Sec.~\ref{sec:feedforward}.

Overall, code shaping does not attempt to dictate the boundaries of user expression or current model capability. Instead, it seeks to classify sketching approaches, highlight tradeoffs, and offer actionable insights for designing interaction. Our study revealed that participants' sketches were highly flexible, adapting to AI performance and specific contexts, making their scope inherently malleable.
For example, in Task 2, some participants used arrows to signify variable type changes, like ``(int, int) $\rightarrow$ (int, string)'', while others annotated function parameters directly. 
Although both text and symbols were interpreted correctly, the model struggled to map between the sketch to the intended edits accurately due to the ambiguity inherent in context-dependent sketches. 
These challenges emphasize the critical role of iteration in code shaping, where users refine their sketches, receive feedforward, and adjust their input to better convey intent.
}

\subsection{Shaping AI Output with Sketches}
While we did not compare sketches to textual prompts directly, some participants (5/18) across the three stages noted that the spatial nature of sketches helped them convey how they wanted to edit \pquote{with more control}{p7}. This suggests a balance between the freedom offered by sketches and the constraints imposed by AI interpretation of code edits. Code shaping tackles this challenge by using freeform sketches to guide the AI interpretation of where and how code edits should be applied, written, performed, or referenced.
Traditional AI-driven code generation tools typically rely on natural language input or UI elements drawn on separate canvases, generating code from different mediums without directly interacting with the code itself. 
This can lead to almost limitless variations in the way natural language is mapped to code structures, which may not always align with the intent of the programmer~\cite{liu_what_2023}.
\rev{One approach exploring the concept of programmable ink, illustrating the potential of combining sketching with computational workflows by enabling users to bind sketches to data and explore outputs dynamically~\cite{inkbase, xia2017writlarge,xia2018dataink, offenwanger2024datagarden}.
However, their focus on binding sketches to predefined computational roles can limit their flexibility for scenarios like code shaping, where the emphasis lies on annotations as interpretative guides rather than functional artifacts. 
Code shaping, therefore, differentiates itself by intentionally keeping sketches free from intrinsic computational meaning but remains the capability to shape AI interpretation by layering sketches on code.}
The combination of sketches and handwritten textual instructions for prompting enhances the precision of the edits while maintaining flexibility~\cite{haught2003creativity}, and balances creative freedom with the necessary structure to achieve desired outcomes.

\subsection{Informal and Formal Programming}
Our findings show that when participants are provided with a pen to code, they approach the program differently (\autoref{sec:conceptual_shift}), 
This approach highlights the contrast between the structured nature of typing code syntax and the more abstract thinking about program structure, flow, and function.
\rev{Code shaping, unlike previous programming-by-example approaches~\cite{10.1145/22627.22349}, extends current programming paradigms by integrating code and sketches, allowing programmers to interact with their work in ways that balance structural precision with creative flexibility (\autoref{fig:classification}). 
This aligns with Olsen’s heuristics \cite{10.1145/1294211.1294256} by demonstrating high expressive leverage and reducing solution viscosity since users can achieve complex edits without articulating structured forms of representation, all while maintaining creative flexibility and structural precision.}

The domain of programming presents a unique opportunity for study, as code takes various shapes highly dependent on its substrate, ranging from editor-based code to syntax within diagrams, visualizations, user interfaces, and even comics~\cite{10.1145/3526113.3545617}.
While there are ongoing discussions comparing differences between text-based programming with higher-level representations~\cite{10.1145/3399715.3399821, noone2018visual}, code shaping aims not to replace typing but to expand the programmer's interaction palette. 
Rather than viewing our research solely as a problem-solving method~\cite{10.1145/3025453.3025765}, we explore new insights and design possibilities emerging from evolving technology~\cite{10.1145/3468505}.
The historical progression from handwritten programs and sketches on printouts to punch cards and eventually typing in an editor illustrates how each programming paradigm unveils unique affordances and constraints~\cite{arawjo_write_2020}. We envision a shared future where programmers can approach their craft through diverse methods, both formal and informal~\cite{pollock2024designing}.
Future research can explore additional representations that bridge the gap between established typed input conventions and the dynamic possibilities of sketch-based interactions, further enriching the programming experience.
\section{Limitations and Future Work}
Our work demonstrates the potential of code shaping as a novel interaction paradigm, but we acknowledge several limitations.
First, while our evaluation utilized Python as the programming language, its flexibility and dynamic nature make it a suitable testbed for prototyping various programming paradigms, including object-oriented, functional, and procedural styles. \rev{Code shaping is not inherently bound to Python or any specific language, as the ink annotations are not tied to computational semantics. While this suggests it might work with other languages, the user experience might differ and required future work to explore how different programming languages potentially influence the effectiveness and usability of sketch-based code editing.}

Second, the current implementation primarily focuses on small codebases (78 lines of code from scenario two), where the relationship between sketches and corresponding code edits is relatively straightforward. \rev{Sketching to edit larger codebases across multiple files might require the implementation of retrieval-augmented generation~\cite{zhang2023repocoder}. Additionally, resolving downstream and upstream implications of code edits, such as propagating variable renames or function refactorings, would require dependency analysis and incremental static analysis techniques to track and update references across the codebase. Currently, these dependencies are implicitly managed by the AI model, but implementing explicit dependency resolution mechanisms, such as abstract syntax tree (AST) traversal or control flow graph (CFG) augmentation, may be necessary for handling larger, interdependent codebases effectively.} This may further involve developing more sophisticated AI models capable of understanding and interpreting complex sketches that span multiple levels of abstraction or integrating visual modeling tools directly within the code editor. Similarly, \rev{while our system supports multiple files as demonstrated in the scenarios, we did not conduct a comprehensive evaluation or support a single ``sketch'' spanning across multiple interdependent files.} Investigating how code shaping can support multi-file editing, maintain context across files, and handle dependencies effectively will be crucial for extending the applicability of this approach to more complex development tasks.
Finally, our study provides initial insights into the potential of code shaping, but further investigation is required to understand its long-term impact on programming practices, particularly in terms of code quality, maintainability, and developer productivity. \rev{We define code shaping in the context of code editing, where sketches are not persistent since they are removed once committed changes are accepted or rejected, or manually deleted. Future research could explore whether versioning sketches is a desirable feature. This could be beneficial for other coding activities such as resolving merge conflicts, refactoring, or asynchronous collaboration.}
\section{Conclusion}
We introduced the concept of code shaping, an interaction paradigm that enables programmers to iteratively edit code using free-form sketch annotations directly on and around the code. Through three stages of design iterations and user studies, we explored how programmers perceive code shaping, the types of sketches they create, common AI interpretation errors, and how they recover from them. We also investigated interface design strategies to effectively bridge the layers between textual code and sketches, such as providing always-on feedforward and integrating unique gestures to minimize context switching. Our findings offer valuable insights into how sketch-based interactions can support code planning and editing, informing future research and design in this emerging area.


\begin{acks}
This work was made possible by 
NSERC Discovery Grant RGPIN-2024-03827, NSERC Discovery Grant \#2020-03966, and Canada foundation for innovation - John R. Evans Leaders Fund (JELF) \#42371.
\end{acks}

\bibliographystyle{ACM-Reference-Format}
\bibliography{_references.bib, shape.bib}


\appendix
\onecolumn

\section{Statistical Result from Three Iterations}

\begin{table}[htbp]
\centering
\small
\begin{tabular}{p{6.8cm}p{.8cm}p{.8cm}p{.8cm}crr}
\toprule
\multirow{2}{*}{\textbf{Metric}} & \multicolumn{3}{c}{\textbf{Median}} & \multicolumn{3}{c}{\textbf{Statistics}} \\
\cline{2-7}
 & \textbf{Iter. 1} & \textbf{Iter. 2} & \textbf{Iter. 3} & \textbf{Comparison} & \textbf{r} & \textbf{p-value} \\
\midrule
\multirow{3}{*}{UMUX-LITE (SUS)} & \multirow{3}{*}{60.82} & \multirow{3}{*}{66.23} & \multirow{3}{*}{82.48} & 1 vs 2 & 1.429 & 0.279 \\
 &  &  &  & 2 vs 3 & 0.000 & 0.066 \\
 &  &  &  & 1 vs 3 & 0.000 & 0.031* \\
\hline
\multirow{3}{*}{NASA-TLX Score} & \multirow{3}{*}{3.92} & \multirow{3}{*}{2.83} & \multirow{3}{*}{1.92} & 1 vs 2 & 2.041 & 0.500 \\
 &  &  &  & 2 vs 3 & 0.817 & 0.138 \\
 &  &  &  & 1 vs 3 & 0.000 & 0.094 \\
\midrule
\multicolumn{7}{c}{\textbf{Self-Defined Likert Scale Questions}} \\
\midrule
\multirow{3}{*}{\makecell[l]{Iterating on my sketches was easy}} & \multirow{3}{*}{4.0} & \multirow{3}{*}{5.0} & \multirow{3}{*}{6.5} & 1 vs 2 & 3.674 & 0.844 \\
 &  &  &  & 2 vs 3 & 0.000 & 0.039* \\
 &  &  &  & 1 vs 3 & 1.429 & 0.156 \\
\hline
\multirow{3}{*}{\makecell[l]{The sketches encapsulated what I intended to achieve}} & \multirow{3}{*}{5.5} & \multirow{3}{*}{5.5} & \multirow{3}{*}{5.5} & 1 vs 2 & 2.654 & 0.783 \\
 &  &  &  & 2 vs 3 & 0.408 & 0.655 \\
 &  &  &  & 1 vs 3 & 1.414 & 0.688 \\
\hline
\multirow{3}{*}{\makecell[l]{The re-generated code aligned with my intended changes}} & \multirow{3}{*}{4.5} & \multirow{3}{*}{5.0} & \multirow{3}{*}{6.0} & 1 vs 2 & 1.225 & 0.892 \\
 &  &  &  & 2 vs 3 & 0.000 & 0.141 \\
 &  &  &  & 1 vs 3 & 0.707 & 0.063 \\
\hline
\multirow{3}{*}{\makecell[l]{I felt more control over the AI model and generated results}} & \multirow{3}{*}{5.5} & \multirow{3}{*}{5.0} & \multirow{3}{*}{5.5} & 1 vs 2 & 2.236 & 0.786 \\
 &  &  &  & 2 vs 3 & 0.000 & 0.785 \\
 &  &  &  & 1 vs 3 & 0.500 & 1.000 \\
\hline
\multirow{3}{*}{\makecell[l]{I felt more control over the whole code editing process}} & \multirow{3}{*}{5.0} & \multirow{3}{*}{5.0} & \multirow{3}{*}{6.0} & 1 vs 2 & 0.000 & 1.000 \\
 &  &  &  & 2 vs 3 & 0.866 & 0.059 \\
 &  &  &  & 1 vs 3 & 1.000 & 0.102 \\
\bottomrule
\end{tabular}
\caption{Study results across three iterations, showcasing the median score, the effect size and p-value for each comparison.}
\label{tab:study_results}
\footnotesize{Note: * indicates p < 0.05}
\end{table}

\section{Types of Action Flow from Stage Three}
\begin{table*}[]
    \centering
    \small
\begin{tabular}{cp{2.2cm}>{\raggedright\arraybackslash}p{4.2cm}>{\raggedright\arraybackslash}p{6cm}c}
\toprule
\textbf{ID} & \textbf{Name} & \textbf{Description} & \textbf{Example Scenario} & \textbf{$N$} \\
\midrule
1 & \textbf{Sketch-Generate-Accept} 
& Participants sketch out code concepts, generate the corresponding code, and accept the generated code without further modifications. 
& P14 sketches a method to sort tasks by due date, generates the code, reviews the output, and accepts it as it correctly implements the sorting logic.
& 32 \\
\cmidrule(lr){1-5}
2 & \textbf{Sketch-Generate-Reject} 
& Participants sketch a concept, generate the code, and then reject the generated code, leading to further refinement of the sketch or code. 
& P18 sketches a Manhattan distance function, generates the code, sees that the distance calculation is incorrect, and revises the sketch by adding further details before generating the code again. 
& 25 \\
\cmidrule(lr){1-5}
3 & \textbf{Cycle2-Sketch} 
& After rejecting the generated code, participants revisit and modify the sketch before generating the code again. 
& P15 rejects the generated task sorting function after realizing it doesn't account for tasks with no due date. The P then revises the sketch to handle missing due dates and regenerates the code. 
& 17 \\
\cmidrule(lr){1-5}
4 & \textbf{Cycle2-Undo/Redo} 
& Participants use undo/redo actions after rejecting generated code, allowing them to adjust their sketches or code without starting from scratch. 
& P13 generates code to impute missing data, but identifies an error in the logic. The P uses the undo function, adjusts the sketch to refine the imputation method, and regenerates the code. 
& 6 \\
\cmidrule(lr){1-5}
5 & \textbf{Cycle2-Edit Code} 
& Participants directly edit the generated code after rejecting the initial output instead of refining the sketch. 
& P17, after rejecting the code generated, decides to manually edit the code for updating task details instead of re-sketching the concept. 
& 2 \\
\cmidrule(lr){1-5}
6 & \textbf{Sketch-Interpret-Generate} 
& Participants sketch an idea, allow the system to interpret it, and then decide whether to accept or reject the system's interpretation and the generated code. 
& P16 sketches a conditional statement. The system interprets the sketch and generates the corresponding code, which the P reviews and accepts as it matches the intended logic. 
& 38/57 \\
\cmidrule(lr){1-5}
7 & \textbf{Sketch-Generate-Compile} 
& Participants generate code from a sketch, compile the code, and then debug and refine the sketch based on compilation or runtime errors. 
& P14 generates the code, compiles it, and encounters a runtime error. The P modifies the sketch to correct the issue, regenerates the code, and recompiles successfully. 
& 11/57 \\
\bottomrule
\end{tabular}
    \caption{Identified action flow types for each iteration from stage three, with example scenarios from the study and frequency counts ($N$).}
    \label{tab:flow}
\end{table*}


\end{document}